\documentclass[journal]{IEEEtran}

\usepackage{todonotes}
\usepackage{listings}
\usepackage{tikz}
\usepackage{comment}
\usepackage{hyperref}
\usetikzlibrary{shapes.geometric, arrows.meta, positioning, decorations.pathreplacing,calc,snakes}
\usetikzlibrary{fit}
\usepackage{amsmath,amssymb}
\usepackage{amsthm}

\definecolor{cadmiumgreen}{rgb}{0.0, 0.42, 0.24}

\definecolor{mygreen}{rgb}{0,0.6,0}
\definecolor{mygray}{rgb}{0.5,0.5,0.5}
\definecolor{mymauve}{rgb}{0.58,0,0.82}

\usepackage{cite}

\ifCLASSINFOpdf
\else
\fi

\usepackage{algorithm}
\usepackage{algpseudocode}

\algnewcommand{\LeftComment}[1]{\Statex \(\triangleright\) #1}
\algnewcommand{\X}{\Statex \(\)}

\usepackage{url}

\begin{document}
%
\title{\emph{Freestyle}, a randomized version of \emph{ChaCha} for resisting offline brute-force and dictionary attacks}
%
%
%

\author{P.~Arun~Babu and
        Jithin~Jose~Thomas

\thanks{P. Arun Babu (arun.babu@rbccps.org) is with the Robert Bosch Center for Cyber-physical Systems, Indian Institute of Science, Bengaluru.}
\thanks{Jithin Jose Thomas (jithint@iisc.ac.in) was with the Department of Electrical Communication Engineering, Indian Institute of Science, Bengaluru.}
}

%
%


\markboth{Manuscript for IEEE Transactions on Information Forensics and Security}%
{Arun \MakeLowercase{\textit{et al.}}: Bare Demo of IEEEtran.cls for IEEE Journals}
%



\maketitle

\begin{abstract}
This paper introduces \emph{Freestyle}, a randomized and variable round version of the ChaCha cipher. Freestyle uses the concept of \emph{hash based halting condition} where a decryption attempt with an incorrect key is likely to take longer time to halt. This makes Freestyle resistant to key-guessing attacks i.e. brute-force and dictionary based attacks. Freestyle demonstrates a novel approach for ciphertext randomization by using random number of rounds for each block, where the exact number of rounds are unknown to the receiver in advance. Freestyle provides the possibility of generating $2^{128}$ different ciphertexts for a given key, nonce, and message; thus resisting key and nonce reuse attacks. Due to its inherent random behavior, Freestyle makes cryptanalysis through known-plaintext, chosen-plaintext, and chosen-ciphertext attacks difficult in practice. On the other hand, Freestyle has costlier cipher initialization process, typically generates 3.125\% larger ciphertext, and was found to be {1.6 to 3.2} times slower than ChaCha20. {Freestyle is suitable for applications that favor ciphertext randomization and resistance to key-guessing and key reuse attacks over performance and ciphertext size. Freestyle is ideal for applications where ciphertext can be assumed to be in full control of an adversary, and an offline key-guessing attack can be carried out.}

\end{abstract}

\begin{IEEEkeywords}
Brute-force resistant ciphers, dictionary based attacks, key-guessing, probabilistic encryption, Freestyle, ChaCha.
\end{IEEEkeywords}

\newcounter{remark}
\newcounter{def}

%
\IEEEpeerreviewmaketitle



\section{Introduction}
\IEEEPARstart{A} randomized (\emph{aka} probabilistic) encryption scheme involves a cipher that uses randomness to generate different ciphertexts for a given $key$, $nonce$ (\emph{a.k.a.} initial vector), and $message$. {The goal of randomization is to make cryptanalysis difficult and a time consuming process}. This paper presents the design and analysis of \emph{Freestyle}, a randomized and variable-round version of ChaCha cipher\cite{bernstein2008chacha}. ChaCha20 (i.e. ChaCha with 20 rounds) is one of the modern, popular (for TLS \cite{langley2016chacha20} and SSH \cite{djm2,djm}), and faster symmetric stream cipher on most machines\cite{chacha-perf,chacha-google-perf}.
Even on lightweight ciphers, realistic brute-force attacks with $key$ sizes $\geq 128$ bits is not feasible with current computational power. However, algorithms and applications that have lower key-space due to: (i) generation of keys from a poor (pseudo-)random number generator \cite{cve-infeon,kim2013predictability,bello2008predictable,yilek2009private,lenstra2012ron,heninger2012mining}; (ii) weak passwords being used to derive keys; and, (iii) poor protocol or cryptographic implementations \cite{vanhoef2017key,beurdouche2015messy,adrian2015imperfect} are prone to key-guessing attacks (brute-force and dictionary based attacks). Also, steady advances are being made in the areas of GPUs \cite{gu2017attacking,agosta2013high,chiriaco2017finding}, specialized hardware for cryptography \cite{wiemer2014high,malvoni2014your,liu2017elliptic,javeed2016high,khalid2017rc4,gurkaynak2017multi}, and  memories in terms of storage and in-memory processing \cite{kim2016multistate, jain2017computing,sebastian2017temporal} to speedup key-guessing attacks.

Techniques such as introducing a delay between incorrect key/password attempts, multi-factor authentication, and CAPTCHAs (Completely Automated Public Turing test to tell Computers and Humans Apart) are being used to resist brute-force attacks over the network (i.e. on-line brute-force attack). However, such techniques cannot be used if the ciphertext is available with the adversary (i.e. offline brute-force attack); for example: encrypted data gathered from a wireless channel, or lost/stolen encrypted files/disks. To resist offline brute-force attacks, key-stretching and slower algorithms\cite{slow-is-good} are preferred. Although, such techniques are useful, they are much slower on low-powered devices, and also slow down genuine users.

This paper makes \emph{three} main contributions: (i) We demonstrate the use of bounded \emph{hash based halting condition}, which makes key-guessing attacks less effective by slowing down the adversary, but remaining relatively computationally simpler for genuine users. We introduce the \emph{key guessing penalty}, which is a measure for a cipher's resistance to key-guessing attacks. The physical significance of KGP is that the adversary would require at least KGP times computational power than a genuine user to launch an effective key-guessing attack; (ii) We demonstrate a novel approach for ciphertext randomization by using random number of rounds for each block of message; where the exact number of rounds are unknown to the receiver in advance; (iii) We introduce the concept of \emph{non-deterministic CTR mode} of operation and demonstrate the possibility of using the random round numbers to generate $2^{128}$ different ciphertexts - even though the $key$, $nonce$, and $message$ are the same. The randomization makes the cipher resistant to key re-installation attacks such as KRACK \cite{vanhoef2017key} and cryptanalysis by XOR of ciphertexts in the event of the $key$ and $nonce$ being reused.

Freestyle attempts to address the following two issues: (i) reuse of a $key$ and $nonce$ combination is not secure in deterministic stream ciphers, as demonstrated attacks such as Key installation attack (KRACK) \cite{vanhoef2017key}. And maintaining a list of used $key$s and $nonce$s is an overhead, especially for constrained and low-powered devices, (ii) Existing ciphers take nearly the same amount of time to decrypt a $message$ irrespective of whether the $key$ used is correct or not. This makes lightweight ciphers prone to key-guessing attacks. The proposed  decryption algorithm in Freestyle is designed to be computationally simpler for a user with a \emph{correct key}; but, for an adversary with an \emph{incorrect key}, the decryption algorithm is likely to take longer time to halt. Thus, each brute-force or dictionary attack attempt is likely to be computationally expensive and time consuming.

\begin{table}
\centering
\caption{\label{tab:notations}List of symbols}
\scriptsize{
\begin{tabular}{|cl|}
\hline
{\bf Notation}&{\bf Description}\\
\hline
{$R_{min}$}&The minimum number of rounds to be used\\
&for encryption. $R_{min} \in [1,2^{16}]$\\\hline
{$R_{max}$}&The maximum number of rounds to be used for\\
&decryption. $R_{max} \in [1,2^{16}]$ and $R_{max} \ge R_{min}$.\\\hline
{$R$}&Number of rounds used to encrypt the current\\
&block of message. {$R$} = ${random}({R_{min}},{R_{max}})$\\\hline
{$R_i$}&Number of rounds used to encrypt $i^{th}$ block of\\
&message. {$R_i$} = ${random}({R_{min}},{R_{max}})$ and $i \ge 0$.\\\hline
{$r$}&The current round number. $r \in [R_{min}, R]$\\\hline
$h()$&Freestyle hash function which generates a 16-bit $hash$.\\\hline
{$H_I$}&Round intervals at which a 16-bit $hash$ has to be\\
&computed. $H_{I} \in [1,R_{min}]$, $R_{min} | H_I$, $R_{max} | H_I$.\\\hline
{$H_C$}&The complexity of Freestyle's hash function \\
&to be used. $H_C \in \{1,2,3\}$ \\\hline
{$I_C$}&The $log_2(iterations)$ (or number of $pepper$ bits)\\
&to be used in during initialization. $I_C \in [8,32]$ \\\hline
$pepper$&The \emph{pepper} value indicating the number of iterations\\
&required during initialization.\\
&$pepper = random (0,2^{{I_C}} - 1)$.\\\hline
{$C_{R_i}$}&The number of rounds computed using an expected\\
&$hash$ and $pepper$ for $i^{th}$ block of $message$.\\\hline
$E_{pepper}$&The expected value of $pepper$.\\\hline
$E_{R_w}$&The expected number of rounds executed by an  \\
&adversary during cipher initialization.\\\hline
$E_R$&The expected number of rounds used by a genuine user \\
&to encrypt/decrypt a block of $message$. If a uniform\\
&distribution is used, then $E_R = \displaystyle\frac{R_{min} + R_{max}}{2}$.\\\hline
{ \color{red}\textbf{\textit{v}}} (in red color)&An input {variable}.\\\hline
{ \color{cadmiumgreen}\textbf{\textit{v}}} (in green color)&A variable derived from one or more input variables.\\\hline
{ \color{blue}\textbf{\textit{v}}} (in blue color)&An output {variable}.\\\hline
{$v^{(r)}$}&The value of {$v$} after {$r$} rounds of Freestyle\\
&If {$v^{(0)}$} is not explicitly defined, then {$v^{(0)} = 0$}.\\\hline
{$v[n]$}&n$^{th}$ element of {$v$}.\\\hline
{$v_1\:||\:v_2$}&Concatenation of $v_1$ and $v_2$.\\\hline
{$v_1\:|\:v_2$}&$v_2$ is a factor of $v_1$.\\\hline
{$v_1\:\oplus\:v_2$}&Bit-wise XOR of $v_1$ and $v_2$.\\\hline
{$v_1\:\boxplus\:v_2$}&Addition of $v_1$ and $v_2$ modulo $2^{32}$.\\\hline
{$v_1\:\boxminus\:v_2$}&Subtraction of $v_1$ and $v_2$ modulo $2^{32}$.\\\hline
{$mod$}&The modulo operator.\\\hline
{$v^*$}&Set of values guessed by an adversary for $v$.\\\hline
{$c_f(v_1,v_2)$}& A set containing common factors of\\
&integers $v_1$ and $v_2$.\\\hline
{$|v|$}&The length of $v$ in bits.\\\hline
$N_b$&The number of blocks in a $message$.\\
&$N_b =\displaystyle\left\lceil\frac{|message|}{512}\right\rceil$\\
&
\\\hline

$Pr_n(X=1)$& The probability of collision of a 16-bit hash\\&at the $n^{th}$ trial when using an incorrect key.\\&\\\hline 
$N_c$&The total number of ciphertexts possible for a given:\\
&$key$, $nonce$, and $message$.\\\hline
$N_r$&The number of ways a block of message can be\\
&encrypted by using random number of rounds.\\
&$N_r = \displaystyle\left(\frac{R_{max} - R_{min}}{H_I} + 1\right)$\\\hline
$T(o)$&The expected time taken to execute the operation $o$.\\\hline
$S$&The 512-bit cipher state for a given block of $message$.\\\hline
$counter$&The counter in CTR mode of operation.\\\hline
$null$&An empty string.\\\hline
\end{tabular}}
\end{table}

\begin{table}[!h]
\centering
\caption{\label{tab:abbreviations}List of abbreviations}
\scriptsize{
\begin{tabular}{|cl|}
\hline
{\bf Abbreviation}&{\bf Expansion}\\
\hline
ARX&Add-Rotate-XOR\\
CAPTCHA&Completely Automated Public Turing test\\
&to tell Computers and Humans Apart\\
CCA&Chosen Ciphertext Attack\\
CPA&Chosen Plaintext Attack\\
CTR&Counter mode of operation\\
DoS&Denial of service\\
HKDF&Halting Key-Derivation Function\\
KGP&Key Guessing Penalty\\
KPA&Known Plaintext Attack\\
KRACK&Key Re-installation Attack\\
MAC&Message Authentication Code\\
MITM&Man In The Middle Attack\\
NONCE&Number used once\\
QR&Quarter Round\\
SSH&Secure Shell\\
TLS&Transport Layer Security\\
\hline
\end{tabular}}
\end{table}

The rest of the paper is structured as follows: Table \ref{tab:notations} and Table \ref{tab:abbreviations} lists the notations and abbreviations used in the paper; section \ref{sec:chacha} presents the background information on ChaCha cipher and its variants; section \ref{sec:freestyle} describes the Freestyle cipher; section \ref{sec:results} presents results and cryptanalysis of Freestyle cipher; section \ref{sec:related-work} presents related work; and section \ref{sec:conclusion} concludes the paper. 

\section{\label{sec:chacha}ChaCha cipher and variants}
ChaCha20\cite{bernstein2008chacha} is a variant of Salsa20\cite{bernstein2005salsa20,bernstein2008salsa20}, a stream cipher. ChaCha20 uses 128-bit $constant$, 256-bit $key$, 64-bit $counter$, and 64-bit $nonce$ to form an initial cipher state denoted by $S^{(0)}$, as:
{\small
\[\begin{bmatrix}
    constant[0],  & constant[1], & constant[2], & constant[3] \\
    \phantom{   xxxx  }key[0],& \phantom{   xxxx  }key[1],&\phantom{   xxxx  }key[2],&\phantom{   xxxx  }key[3] \\
    \phantom{   xxxx  }key[4],& \phantom{   xxxx  }key[5],&\phantom{   xxxx  }key[6],&\phantom{   xxxx  }key[7] \\
    \phantom{x}counter[0],&\phantom{x}counter[1],&\phantom{xx}nonce[0],&\phantom{xx}nonce[1] \\
\end{bmatrix}
\]}

ChaCha20 uses 10 double-rounds (or 20 rounds) on $S^{(0)}$; where each of the double-round consists of 8 quarter rounds(QR) defined as:

\begin{equation}\label{eq:odd-rounds}
{\small
\begin{array}{cccc}
       QR\:\:(S[0], &S[4], &S[\phantom{0}8], &S[12])\\
       QR\:\:(S[1], &S[5], &S[\phantom{0}9], &S[13])\\
       QR\:\:(S[2], &S[6], &S[10], &S[14])\\
       QR\:\:(S[3], &S[7], &S[11], &S[15])\\
\end{array}}
\end{equation}
\begin{equation}\label{eq:even-rounds}
{\small
\begin{array}{cccc}
       QR\:\:(S[0], &S[5], &S[10], &S[15])\\
       QR\:\:(S[1], &S[6], &S[11], &S[12])\\
       QR\:\:(S[2], &S[7], &S[\phantom{0}8], &S[13])\\
       QR\:\:(S[3], &S[4], &S[\phantom{0}9], &S[14])
\end{array}}
\end{equation}

The 16 elements of the cipher-state matrix are denoted by using an index in range [0,15], and the quarter-round $QR(a,b,c,d)$ is defined as:
\begin{equation}
 \begin{array}{ccc}\label{eq:qr}
    a \leftarrow a \boxplus b; & d \leftarrow d \oplus a; &d \leftarrow d \lll 16;\\
    c \leftarrow c \boxplus d; & b \leftarrow b \oplus c; &b \leftarrow b \lll 12;\\
    a \leftarrow a \boxplus b; & d \leftarrow d \oplus a; &d \leftarrow d \lll \phantom{1}8;\\
    c \leftarrow c \boxplus d; & b \leftarrow b \oplus c; &b \leftarrow b \lll \phantom{1}7;\\
  \end{array}
\end{equation}

After 20 rounds, the initial state $(S^{(0)})$ is added to the current state $(S^{(20)})$ to generate the final state. The final state is serialized in the little-endian format to form the 512-bit key-stream, which is then XOR-ed with a block (512 bits) of plaintext/ciphertext to generate a block of ciphertext/plaintext. The above operations are performed for each block of $message$ to be encrypted/decrypted.

ChaCha is a simple and efficient ARX (Add-Rotate-XOR) cipher, and is not sensitive to timing attacks. ChaCha has two main flavors with reduced number of rounds i.e. with 8 and 12 rounds. ChaCha8 is considered secure enough as there are no known attacks against it yet. ChaCha20 has two main variants: (i) IETF's version of ChaCha20 \cite{langley2016chacha20,nir2015chacha20} which uses a 32-bit $counter$ (instead of 64-bit) and 96-bit $nonce$ (instead of 64-bit); and (ii) XChaCha20 \cite{denisxchacha20}, which uses 192-bit $nonce$ (instead of 64-bit), where a randomly generated $nonce$ is considered safe enough \cite{libsodium-report}. The large $nonce$ in XChaCha20 makes the probability of $nonce$ reuse low.

\section{\label{sec:freestyle}The \emph{Freestyle} cipher}
\subsection{\label{sec:hbsc}Hash based halting condition}
Traditionally ciphers are designed to use fixed number of rounds in the encryption and decryption process. This makes the cipher to take nearly the same amount of time to execute the decryption function irrespective of the $key$ being correct or incorrect. This is advantageous for an adversary if the cipher is lightweight and parallelizable. To resist such attacks, we use the concept of \emph{hash based halting condition}.

The purpose of hash based halting condition is to make decryption take longer time to halt if an incorrect $key$ is used. It works on the principle that the exact number of rounds to decrypt a block of $message$ is not shared with the receiver, but can be computed by the receiver using the correct $key$ and one or more $hash$es. The $hash$es must be shared by the sender in cleartext along with the ciphertext. The number of rounds ($R$) to be used to encrypt a given block is  generated randomly by the sender from the range $[R_{min}, R_{max}]$; and only an expected $hash$ of the state of the cipher after running $R$ rounds are shared. The expected $hash$ acts as a stop condition for decrypting a block of message; and the receiver has to execute the decryption algorithm till the computed $hash$ equals the expected $hash$. For an adversary using brute-force or dictionary based attack, since the $key$ is incorrect, during the decryption process the {$hash$} is expected to take longer time to match (with high probability). This property makes offline brute-force and dictionary based attacks less efficient. The hash based halting condition is only applicable to ciphers having a symmetric structure (e.g. use of feistel network).

\stepcounter{remark}
\noindent{\bf Remark \arabic{remark} } For better security, $R$ must be generated using a good uniform random number generator like hardware random number generator or cryptographically secure pseudo-random number generator (e.g. {\tt arc4random} \cite{arc4random}).

\stepcounter{remark}
\noindent{\bf Remark \arabic{remark} } The proposed approach makes the assumption that the hash function is secure enough, that from the $hash$ it is computationally infeasible to compute the number of rounds, $key$, or any other secret information.

\subsection{\label{sec:cipher-config}Cipher parameter}

The Freestyle cipher is formally defined as $Freestyle({R_{min}}, {R_{max}}, {H_C}, H_I, I_C)$; where { $R_{min}$}, { $R_{max}$} indicate the minimum and maximum number of rounds to be used for encryption/decryption respectively. {$H_C \in \{1,2,3\}$, indicates the level of complexity of hash function to be used; where 1 indicates the lowest complexity, the highest performance, and the lowest security; and 3 indicates the highest complexity, the lowest performance, and the highest security.  {$H_C$} is also used to determine the number of quarter rounds (QRs) to be used to compute the $hash$. $H_I$ indicates the round intervals at which a 16-bit {$hash$} of cipher-state must be computed. And $I_C \in [8,32]$ indicates the number of bits used to generate a random number ($pepper$) which is chosen between $[0,2^{I_C})$. The $pepper$ value is used as number of iterations performed to initialize the cipher. The $pepper$ in general is a number which has the same function as $salt$, but is usually of fewer bits, and is not stored along with the hash or ciphertext (i.e. can be forgotten by the sender after use) \cite{kedem1999bruteforceattackonuni,forler2013catena}.
At initialization, Freestyle concatenates {$R_{min}$}, { $R_{max}$}, {$H_C$}, and {$H_I$}; to generate a unique 64-bit {$cipher\_parameter$} as shown in the figure \ref{fig:config}.
\begin{figure}[!h]
\centering
{\small
\begin{tabular}{|c|c|c|c|c|}
\hline
$R_{min}$&$R_{max}$&$H_{I}$&$H_{C}$&$I_{C}$\\
(16-bits)&(16-bits)&(16-bits)&(8-bits)&(8-bits)\\
\hline
\end{tabular}}
\caption{\label{fig:config} The 64-bit $cipher\_parameter$}
\end{figure}

The {$cipher\_parameter$} is to be XOR-ed with the $key$ (equation \ref{eq:initial-state}), which makes encryption with one {$cipher\_parameter$} incompatible with other cipher parameters by design; thus cryptanalysis data collected for a weaker {$cipher\_parameter$} cannot be used directly for other parameters. For a given $cipher\_parameter$, the total number of ways a block of $message$ can be encrypted using random number of rounds (in the range $[R_{min},R_{max}]$) which is denoted by $N_r$, given as:
\begin{equation}\label{eq:total}
N_r = \displaystyle\frac{R_{max} - R_{min}}{H_I} + 1\\
\end{equation}

\stepcounter{remark}
\noindent{\bf Remark \arabic{remark} }While choosing a $cipher\_parameter$, it must be noted that the performance of Freestyle is $\propto \frac{H_I}{H_C\times I_C}$. The value of $R_{min}$ must be chosen carefully based on the required security level, and is recommended that $R_{min}$ be at least 8 as there are no known attacks for ChaCha8. For security-critical applications though, $R_{min} \ge 12$ is preferred. To have better randomization, it is recommend that $N_r \geq 4$; also, as there are only $2^{16}$ unique possible $hash$es represented by a 16-bit unsigned integer; $R_{min}, R_{max},$ and $H_I$ must be chosen such that the following relationship holds (from equation \ref{eq:total}):
\begin{equation}\label{eq:recommend}
\begin{array}{ccccc}
3&\leq&\displaystyle\frac{R_{max} - R_{min}}{H_I}&\leq&65535
\end{array}
\end{equation}
Also, for better security, the recommended values for $H_C$ is 3 or 2, and for $H_I$ it is 1 or 2. $I_C$ must be chosen based on performance and the security level required, and $I_C \ge 20$ is recommended for security-critical applications.\hfill$\blacksquare$

\subsection{\label{sec:matrix}The initial cipher state}
The initial cipher state of Freestyle, denoted by $S^{(0)}$ (equation \ref{eq:initial-state}) is a 4$\times$4 matrix of 32-bit words consisting of 128-bit $constant$, 256-bit $key$, 32-bit $counter$, and 96-bit $nonce$. Unlike ChaCha, the $counter$ size has been reduced to 32-bit as in practice most of the protocols such as the SSH transport protocol \cite{ylonen2006secure} recommend re-keying after 1GB of data sent/received. 

The initial cipher state acts the input for generating a key-stream for a block of $message$. The cipher state of Freestyle is similar to the IETF's version of {ChaCha}, except that the $constant$, $key$, and $counter$ is modified as shown in equation \ref{eq:initial-state}. The initial-state has been modified in such a way that: either a publicly known value is XOR-ed with a secret element of the matrix, or a secret value is XOR-ed with a publicly known element of the matrix.

Here, we introduce the \emph{non-deterministic CTR mode} of operation where, the $counter$ is XOR-ed with a random value that is independent of the $key$ or $nonce$ (unlike randomized-CTR mode where the random number is derived from $key$ and/or $nonce$). Hence, the property of CTR mode of operation: that the difference between  the $counter$s of $(n+1)^{th}$ block and $n^{th}$ block is equal to 1 may no longer hold. The random number to be XOR-ed with $counter$ in Freestyle is denoted by $random\_word[3]$, and its value is initialized during cipher initialization (section \ref{sec:random-init} and figure \ref{fig:random}). The Freestyle cipher starts with the plain CTR mode of operation and shifts to \emph{non-deterministic CTR mode} after 28 blocks (i.e. after random number initialization), thus making cryptanalysis difficult.

\begin{figure*}[!ht]
 \line(1,0){515} 
{\small
\begin{equation}\label{eq:initial-state}
S^{(0)} = \left[\begin{array}{cccc}
    \left(\begin{array}{c}constant[0] \\\oplus \\random\_word[0]\end{array}\right),  &\left(\begin{array}{c}constant[1] \\\oplus \\random\_word[1]\end{array}\right), & \left(\begin{array}{c}constant[2] \\\oplus \\random\_word[2]\end{array}\right), &  constant[3] \\
    
\left(\begin{array}{c}key[0] \\\oplus \\cipher\_parameter[0]\end{array}\right),  & \left(\begin{array}{c}key[1] \\\oplus \\cipher\_parameter[1]\end{array}\right) , &  key[2], & key[3]\\
&&&\\
    key[4],  & key[5], & key[6], & key[7]\\
    &&&\\
    \left(\begin{array}{c}counter \\\oplus \\random\_word[3]\end{array}\right),  & nonce[0], & nonce[1], & nonce[2]
\end{array}\right]
\end{equation}}
 \line(1,0){515}
\end{figure*}

\begin{figure*}
\centering
\includegraphics{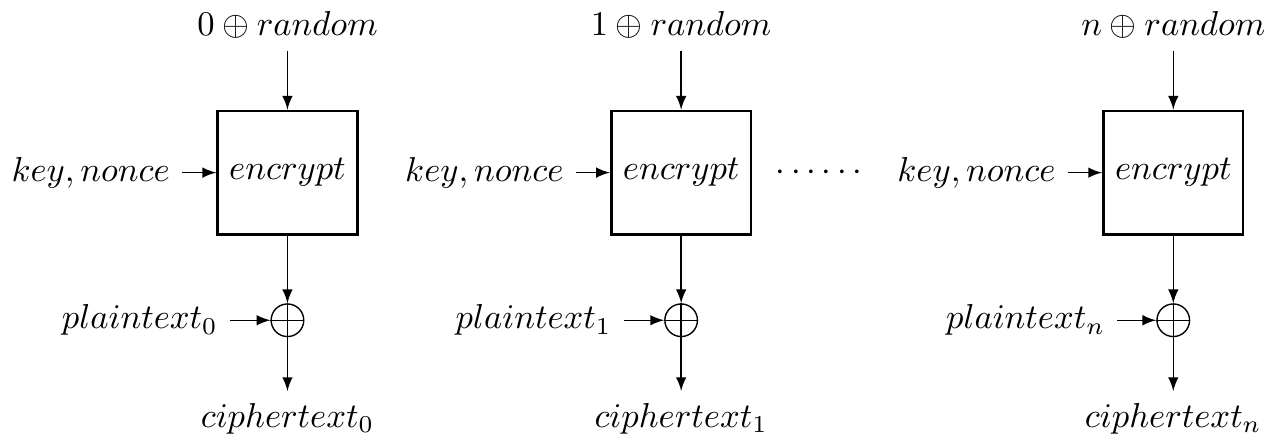}
\caption{\label{fig:rctr}Non-deterministic CTR mode of operation, where the $counter$ is XOR-ed with a random number that is independent of the $key$ and $nonce$}
\end{figure*}

In equation \ref{eq:initial-state}, the $random\_word$s indicate the 128-bit random number generated by the sender, which can be computed by the receiver using the correct $key$. The $random\_word$s are initially set to 0 by both sender and receiver and must be computed while initializing the cipher (section \ref{sec:random-init}). 
Using the initial cipher state, Freestyle uses ChaCha's approach to generate the final state (equations \ref{eq:odd-rounds}, \ref{eq:even-rounds}, \ref{eq:qr}); however unlike ChaCha, Freestyle supports both even and odd number of rounds.

\subsection{Hash function}
\begin{figure}
\centering
\includegraphics[scale=0.75]{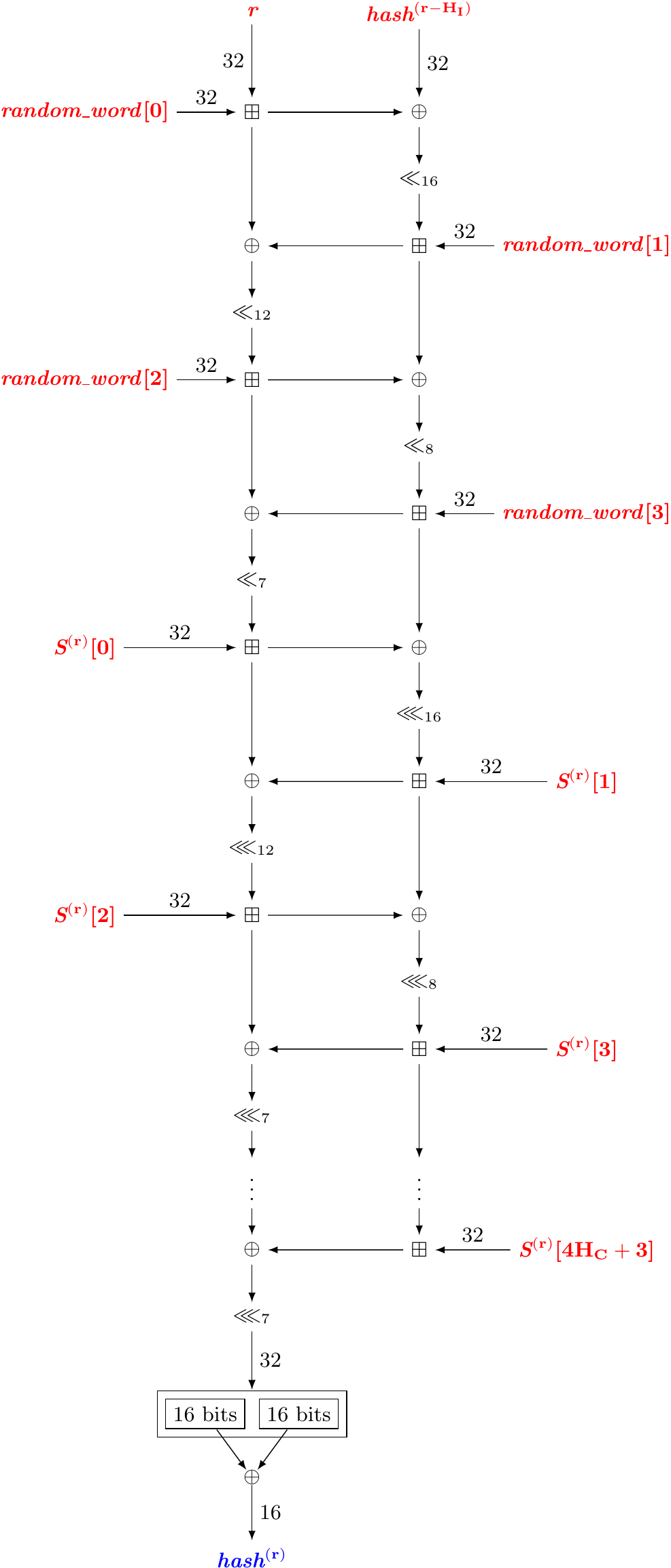}
\caption{\label{fig:hash}The Freestyle hash function - $h(
)$, for the round $r$ (the size of variables are in bits). Note that the value of $hash^{(R_{min}-H_I)}$ is always 0.}
\end{figure}

Freestyle's hash function is used to generate the hash based halting condition described in section \ref{sec:hbsc}. The hash function (figure \ref{fig:hash}) generates a 16-bit $hash$ using: (i) the current round number ({$r$}), (ii) the first $128\times(H_C + 1)$ bits of current cipher state ($S^{(r)}$), (iii) the 128-bit $random\_word$s, and (iv) the previous $hash$ (i.e. $hash^{(r-H_I)}$). 

To resist timing-based or side-channel attacks, the hash function uses Add-Rotate-XOR (ARX) operations, the same set of operations used by Freestyle quarter-round (QR). Also, unlike a typical cryptographic hash function, Freestyle does not require high collision-resistant hash function. The probability of $\frac{1}{2^{16}}$ for collision is enough for its purpose.} 

\subsection{\label{sec:random-init}Random number initialization}

\begin{figure}\centering
\includegraphics[scale=0.75]{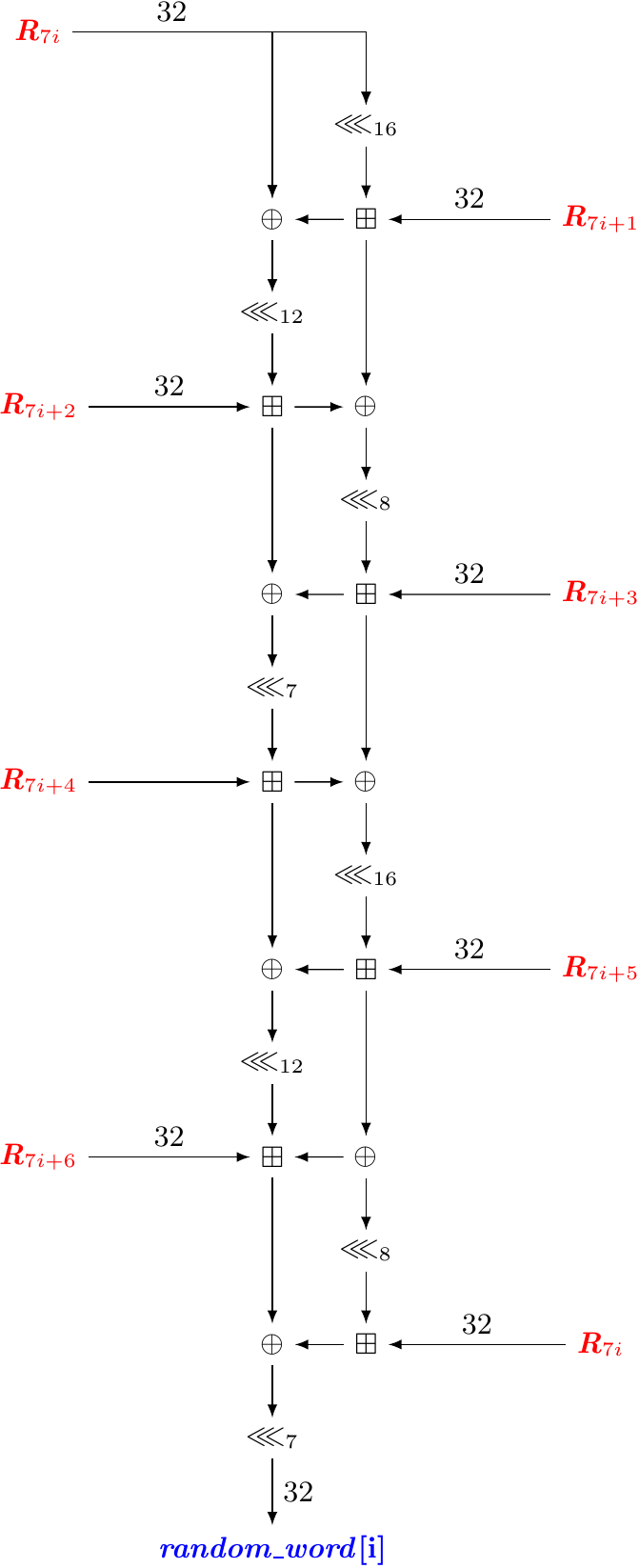}
\caption{\label{fig:random}
Generation of $random\_word[i]$, where $i \in [0,3]$ (the size of variables are in bits)}
\end{figure}

As mentioned earlier in section \ref{sec:hbsc}, Freestyle uses random number of rounds to encrypt a $message$ (equation \ref{eq:total}). To randomize ciphertext even further, Freestyle requires the sender to generate a 128-bit random number denoted by $random\_word$s; that will act as one of the inputs for encryption and decryption. Freestyle  enables a sender to securely send $random\_word$s to the receiver even though the $key$ and $nonce$ may be reused. 

After the $cipher\_parameter$ is computed (section \ref{sec:cipher-config}), the following temporary configuration is set irrespective of the $cipher\_parameter$: \begin{align}R_{min} = 12,\: R_{max} = 36,\: H_C = 3,\: H_I = 1\end{align}
This is done to ensure there is enough entropy even if weaker values of $R_{min}$ and $R_{max}$ are provided by the user; and also in cases where the parameters can be downgraded in Man in the middle (MITM) attacks such as Logjam \cite{adrian2015imperfect}.

The sender then sets $random\_word$s to 0 and generates a random $pepper$ ($p$) in the range $[0,2^{I_C})$, which is added to the initial cipher-state. The sender then generates 28 random numbers ($R_{0}$ to $R_{27}$) in the range $[12,36]$ using a uniform distribution. Each of the 28 random numbers is then used as number of rounds (equations \ref{eq:odd-rounds},\ref{eq:even-rounds}, and  \ref{eq:qr}) in Freestyle cipher to generate 28 $hash$es after executing $R_i$ rounds, where $i \in [0,27]$ (figure \ref{fig:hash}). It must be noted that for each of the 28 round numbers, no encryption is performed, only expected $hash$es are generated. The sender also ensures that hash collisions are handled correctly, which is a crucial step for correct decryption by the receiver. The sender then sends the 28 $hash$es to the receiver, and computes $random\_word$s from $R_i$ values as shown in figure \ref{fig:random}.

On the other hand, the receiver first sets the $random\_word$s to 0; and increments $S^{(0)}[3]$ (i.e. the $constant[3]$) and for each increment, computes 28 $hash$es, until the computed $hash$es equals with the received 28 $hashe$s. Receiver then computes the $R_0$ to $R_{27}$ from: $key$, $nonce$, and 28 $hash$es. Using which, $random\_word$s are computed as shown in figure \ref{fig:random}.

Finally, both sender and receiver will: reset the $counter$ to 0, set $R_{min}$, $R_{max}$, $H_C$, and $H_I$ to their original values, and the new initial cipher state is computed using equation \ref{eq:initial-state}. From now on, the new initial cipher state will be used for encryption/decryption (section \ref{sec:encryption}). The steps to initialize the Freestyle cipher are described in Algorithm \ref{algo:init-sender} and \ref{algo:init-receiver}.

\begin{algorithm}
\caption{Freestyle initialization for the sender}\label{algo:init-sender}
\begin{algorithmic}[1]
\Procedure{freestyle\_init\_sender}{}
\X{{\bf Inputs:} $S^{(0)}, R_{min}, R_{max}, H_I, H_C$}
\X{}

\State Save the values of $R_{min}, R_{max}, H_C,$ and $H_I$
\State Set $R_{min}\gets 12$, $R_{max}\gets 36$, $H_{C}\gets 3$, $H_{I}\gets 1$
\State Set $random\_word[i] \gets 0,\: \forall i \in [0,3]$

\State $pepper \gets random(0,2^{I_C} - 1)$

\State $S^{(0)}[3] \gets S^{(0)}[3] \boxplus pepper$

\X{}

\LeftComment{Generate 28 hashes using 28 random number of rounds}
\For{$i\gets 0$ to $ 27$}
\State $\{R_i,hash[i]\} \gets freestyle\_encrypt\_block\: ($
\X{\qquad\qquad\qquad\qquad\qquad\qquad$S^{(0)},$}
\X{\qquad\qquad\qquad\qquad\qquad\qquad$null,$}
\X{\qquad\qquad\qquad\qquad\qquad\qquad$random\_word,$}
\X{\qquad\qquad\qquad\qquad\qquad\qquad$R_{min},$}
\X{\qquad\qquad\qquad\qquad\qquad\qquad$R_{max},$}
\X{\qquad\qquad\qquad\qquad\qquad\qquad$H_{I},$}
\X{\qquad\qquad\qquad\qquad\qquad\qquad$H_{C},$}
\X{\qquad\qquad\qquad\qquad\qquad\qquad$i$}\Comment{the counter}
\X{\qquad\qquad\qquad\qquad\qquad)}
\EndFor
\X{}
\LeftComment{Check if the receiver will find a hash collision between
0 and $(pepper - 1)$. If yes, update $R_i, \forall i \in [0,27]$}
\State $S^{(0)}[3] \gets S^{(0)}[3] \boxminus pepper$ \Comment{Restore constant}
\For{$p\gets 0$ to $(pepper - 1)$}
\For{$i\gets 0$ to $ 27$}
\State $C_{R_i} \gets freestyle\_decrypt\_block\: ($
\X{\qquad\qquad\qquad\qquad\qquad$S^{(0)},$}
\X{\qquad\qquad\qquad\qquad\qquad$null,$}
\X{\qquad\qquad\qquad\qquad\qquad$hash[i],$} \Comment{expected hash}
\X{\qquad\qquad\qquad\qquad\qquad$random\_word,$}
\X{\qquad\qquad\qquad\qquad\qquad$R_{min},$}
\X{\qquad\qquad\qquad\qquad\qquad$R_{max},$}
\X{\qquad\qquad\qquad\qquad\qquad$H_{I},$}
\X{\qquad\qquad\qquad\qquad\qquad$H_{C},$}
\X{\qquad\qquad\qquad\qquad\qquad$i$}\Comment{the counter}
\X{\qquad\qquad\qquad\qquad)}
\If{$C_{R_i} = 0$}
\State {\bf goto} step 20 \Comment{Increment $pepper$ and retry}
\EndIf

\EndFor
\X{}
\State $R_i \gets C_{R_i},\:\forall i \in [0,27]$ \Comment{Found a collision}
\State {\bf break}
\X{}
\State $S^{(0)}[3] \gets S^{(0)}[3] \boxplus 1$ \Comment{Retry}
\EndFor

\X{}

\State Compute $random\_words$ (as given in figure \ref{fig:random})
\State Restore the original values of $R_{min}, R_{max}, H_C, H_I$
\State $S^{(0)}[12] \gets 0$ \Comment{Reset counter}
\X{}

\State \textbf{return} $hash[i],\:\forall i \in[0,27] $
\EndProcedure
\end{algorithmic}
\end{algorithm}

\begin{algorithm}
\caption{Freestyle initialization for the receiver}\label{algo:init-receiver}
\begin{algorithmic}[1]
\Procedure{freestyle\_init\_receiver }{}
\X{{\bf Inputs:} $S^{(0)}, cipher\_parameter, hash$}
\X{}
\State Save the values of $R_{min}, R_{max}, H_C,$ and $H_I$
\State Set $R_{min}\gets 12$, $R_{max}\gets 36$, $H_{C}\gets 3$, $H_{I}\gets 1$
\State Set $random\_word[i] \gets 0,\: \forall i \in [0,3]$

\X{}

\For{$pepper \gets 0$ to $(2^{I_C} - 1)$}
\For{$i \gets 0$ to $27$}
\State $C_{R_i} \gets freestyle\_decrypt\_block\: ($
\X{\qquad\qquad\qquad\qquad\qquad$S^{(0)},$}
\X{\qquad\qquad\qquad\qquad\qquad$null,$}
\X{\qquad\qquad\qquad\qquad\qquad$hash[i],$} \Comment{expected hash}
\X{\qquad\qquad\qquad\qquad\qquad$random\_word,$}
\X{\qquad\qquad\qquad\qquad\qquad$R_{min},$}
\X{\qquad\qquad\qquad\qquad\qquad$R_{max},$}
\X{\qquad\qquad\qquad\qquad\qquad$H_{I},$}
\X{\qquad\qquad\qquad\qquad\qquad$H_{C},$}
\X{\qquad\qquad\qquad\qquad\qquad$i$} \Comment{the counter}
\X{\qquad\qquad\qquad\qquad)}
\If{$C_{R_i} = 0$}
 \State {\bf goto} step 13 \Comment{Increment $pepper$ and retry}
\EndIf
\EndFor
\X{}
\State {\bf break} \Comment{Found all 28 valid round numbers ($R_i$)}
\X{}
\State $S^{(0)}[3] \gets S^{(0)}[3] \boxplus 1$ \Comment{Retry}
\EndFor

\X{}

\State Compute $random\_words$ (as given in figure \ref{fig:random})
\X{}
\State Restore the original values of $R_{min}, R_{max}, H_C, H_I$

\State $S^{(0)}[12] \gets 0$ \Comment{Reset counter}

\X{}

\EndProcedure
\end{algorithmic}
\end{algorithm}

\begin{algorithm}
\caption{Encryption of a block of message}\label{algo:encryption}
\begin{algorithmic}[1]
\Procedure{freestyle\_encrypt\_block  }{}
\X{{\bf Inputs:} $S^{(0)}, plaintext, random\_word,$}
\X{\qquad\qquad$R_{min}, R_{max}, H_I, H_C, counter$}
\X{}
\State $hash \gets 0$
\X{}
\State {$collided [h] \gets false,\: \forall h \in [0,2^{16})$}
\X{}
\State $S^{(0)}[12] \gets counter \oplus random\_word[3]$
\X{}
\State $R \gets random(R_{min}, R_{max})$
\X{}
\For{$r \gets 1$ to $R$}
\X{}
\State Compute $S^{(r)}$ using Freestyle (Equations \ref{eq:odd-rounds}, \ref{eq:even-rounds})
\X{}
\If{$r \geq R_{min}$ {\bf and} $r | H_I$}
\State $hash \gets h(S^{(r)}, r, random\_words, hash)$
\While {$collided [hash] = true$}
\State $hash \gets (hash + 1)\:mod\:(2^{16})$
\EndWhile
\State $collided[hash] = true$
\EndIf
\X{}
\EndFor
\X{}
\If{$plaintext = null$} \Comment{While initialization}
\State \textbf{return} \{$R$, $hash$\}
\Else
\State $keystream \gets little\_endian (S^{(R)} \boxplus S^{(0)})$
\State $ciphertext \gets plaintext \oplus keystream$

\X{}
\State \textbf{return} \{$R$, $hash$, $ciphertext$\}

\EndIf
\EndProcedure
\end{algorithmic}
\end{algorithm}
\begin{algorithm}
\caption{Decryption of a block of message}\label{algo:decryption}
\begin{algorithmic}[1]
\Procedure{freestyle\_decrypt\_block}{}
\X{{\bf Inputs:} $S^{(0)}, plaintext, expected\_hash, random\_word,$}
\X{\qquad\qquad$R_{min}, R_{max}, H_I, H_C, counter$}
\X{}
\State $R \gets 0$
\State $hash \gets 0$
\X{}
\State {$collided [h] \gets false,\: \forall h \in [0,2^{16})$}
\X{}
\State $S^{(0)}[12] \gets counter \oplus random\_word[3]$
\X{}
\For{$r \gets 1$ to $R_{max}$}
\X{}
\State Compute $S^{(r)}$ using Freestyle (Equations \ref{eq:odd-rounds}, \ref{eq:even-rounds})
\X{}
\If{$r \geq R_{min}$ {\bf and} $r | H_I$}
\State $hash \gets h(S^{(r)}, r, random\_words, hash)$
\While {$collided [hash] = true$}
\State $hash \gets (hash + 1)\:mod\:(2^{16})$
\EndWhile
\X{}
\If{$hash = expected\_hash$}
\State $R \gets r$
\State {\bf break}
\EndIf
\X{}
\State $collided[hash] = true$
\EndIf
\X{}
\EndFor
\X{}
\If{$plaintext = null$} \Comment{While initialization}
\State \textbf{return} $R$
\Else

\State $keystream \gets little\_endian (S^{(R)} \boxplus S^{(0)})$
\State $plaintext \gets ciphertext \oplus keystream$

\X{}
\State \textbf{return} \{$R$, $plaintext$\}

\EndIf
\EndProcedure
\end{algorithmic}
\end{algorithm}
\stepcounter{remark}
\noindent{\bf Remark \arabic{remark} } 
The rationale behind using 28 random number of rounds to generate a 128-bit $random\_word$s is: as the total possible random numbers that a sender can choose between $[12,36]$ using $H_I = 1$ is 25 (equation \ref{eq:total}).
 Thus, if the sender has to generate $n$ random numbers, and as the $random\_word$s is a 128-bit value, $25^{n} \ge 2^{128}$ to be a good random number generator. Thus, $n$ must be at least 28. The $random\_word$s provide the possibility of generating $2^{128}$ different ciphertexts for a given $key$, $nonce$, and $message$.\\
 
 \stepcounter{remark}
\noindent{\bf Remark \arabic{remark} } 
The proposed approach is different from generating a 128-bit random number ($\mathcal{R}$) and sending it in encrypted form. For example: 
\begin{align}
encrypt (\mathcal{R}, key)\:||\:encrypt^{'}(message, key, \mathcal{R})
\end{align} In the latter case, for stream-ciphers, if the $key$ and $nonce$ are reused, there is a possibility of cryptanalysis by XOR-ing ciphertexts.

\subsection{\label{sec:encryption}Encryption and decryption}
After the computation of $random\_word$s and the new initial cipher state ($S^{(0)}$); to encrypt a block of $message$, the sender generates a random number ({$R$}) in the range {$[R_{min}$}, {$R_{max}]$}, using which a key-stream and a $hash$ are generated after {$R$} rounds of Freestyle.
The plaintext is XOR-ed with the key-stream to generate the ciphertext. The ciphertext along with the expected $hash$ is sent to the receiver.

On the other hand, to decrypt a block of $message$, the receiver computes the $S^{(r)}$ using $n$ number of $H_I$ rounds of Freestyle until $R_{max}$ rounds or until the computed {$hash$} at the end of each $H_I$ rounds equals with the received {$hash$}. After which, a key-stream is generated which is then XOR-ed with the ciphertext to generate the plaintext. The steps to encrypt/decrypt a block (512 bits) of $message$ is described in Algorithm \ref{algo:encryption} and \ref{algo:decryption}, and are to be performed for each block of $message$ to be exchanged.\\

\stepcounter{remark}
\noindent{\bf Remark \arabic{remark} }
If both sender and receiver have used the same {$cipher\_parameter$}, $random\_word$s, {$key$},  and {$nonce$}; then the sender and receiver would have taken same number of steps and operations (i.e. {$R$} rounds) to generate the key-stream for a given block of $message$. 

\stepcounter{remark}
\noindent{\bf Remark \arabic{remark} }
During initialization, Freestyle's hash function uses 512 bits of $S^{(R)}$ and 16 bit value of current round number $r$ as inputs to generate a 16-bit $hash$. Whereas during encryption/decryption the hash function uses at least 256-bits of $S^{(R)}$ and 128-bit $random\_word$s. It is computationally infeasible to compute the $key$ or key-stream using brute-force approach, as it would require at least $2^{320}$ {operations} (i.e. 256 bits of $S^{(R)}$ and at least 64 bits of Add-Rotate-XOR result of $r, hash^{(r-H_I)},$ and $random\_words$) to generate all possible cipher states (or partial cipher-states in case of encryption/decryption) that may collide with a given $hash$ (figure \ref{fig:hash}). Also, assuming the 16-bit $hash$es are equally spread over $2^{16}$ buckets, there are likely to be $2^{304}$ collisions.

\section{\label{sec:results}Results and discussions}

\subsection{\label{sec:num-possible-ciphers}Number of possible ciphertexts}

For a given $message$ of length $|message|$ bits, the $message$ is divided into $N_b = \left\lceil\frac{|message|}{512}\right\rceil$ blocks. Since, each block can be encrypted with a random number ($R$) of rounds in the range $[R_{min}, R_{max}]$. And since all the blocks of the $message$ use the 128-bit $random\_word$s as input, the total number of possible ciphertexts are: 
\begin{equation} \label{eq2}
\begin{split}
N_c\:=&\:2^{128}\times (N_r)^{\displaystyle N_b}\\
\end{split}
\end{equation}From equation \ref{eq2}, as the number of blocks in a $message$ increases, the number of possible ciphertext increases exponentially.

\subsection{Resisting cryptanalysis}

\subsubsection{\label{sec:cryptanalysis}Known-plaintext attacks (KPA), Chosen-plaintext attacks (CPA), and differential cryptanalysis}
For a known or chosen plaintext, due to the random behavior of Freestyle, even if the $nonce$ is controlled by the adversary, there are $N_c$ possible ciphertexts. Hence, the effort required in cryptanalysis using known plaintext, chosen plaintext, differential analysis increases $N_c$ times.

\subsubsection{Chosen-ciphertext attacks (CCA)}

In chosen-ciphertext attacks we consider two cases based on the adversary's ability to control the $nonce$.

\paragraph{If $nonce$ cannot be controlled by the adversary}

To generate a ciphertext, an adversary while initializing the cipher (section \ref{sec:random-init}) has to provide 28 valid $hash$es, and at least one valid $hash$ for sending block(s) of ciphertext. As a random round is chosen between [12,36] to initialize the $random\_word$s (equation \ref{eq:total}), there are only 25 valid values for $hash$. While performing decryption, the total possible $hash$es that can be accepted by the receiver for a block of ciphertext is $N_r = \left(\frac{R_{max} - R_{min}}{H_I} + 1\right)$. And as there are $2^{16}$ possible values for $hash$, to send a valid ciphertext, the adversary has to send (28 + $N_b$) valid $hash$es. By brute-force approach, the probability of such an event occurring is:

\begin{align}
\displaystyle\left(\frac{25}{2^{16}}\right)^{28} \times \left( \frac{N_r}{2^{16}}\right)^{N_b}\\
\nonumber\\
< \displaystyle\frac{1}{2^{317}}
\end{align}

Assuming a constant time cryptographic implementation to check the validity of (28 + $N_b$) $hash$es, it is infeasible to generate a ciphertext that can be accepted by a receiver. This makes chosen-ciphertext attacks difficult in practice if $nonce$ cannot be controlled by the adversary. 

\paragraph{If $nonce$ can be controlled by the adversary}

In this case, the adversary can launch CPA which can reveal (28 + $N_b$) valid $hash$es. Thus, the adversary can replay them to make the receiver accept arbitrary ciphertext of $N_b$ blocks.

In either of the two cases, after successfully sending a valid ciphertext, the adversary still has to guess the 128-bit $random\_word$s. It is computationally infeasible to know which combination of $key$ and $random\_word$s the 28 $hash$es map to.

\stepcounter{remark}
\noindent{\bf Remark \arabic{remark} }
It must be noted that Freestyle's hash function does not use $message$ as an input. Hence, cannot prevent ciphertext tampering. In practice, Freestyle like ChaCha must be used with a message authentication code (MAC) such as Poly1305 \cite{bernstein2005poly1305}.

\subsubsection{XOR of ciphertexts when $key$ and $nonce$ are reused}
Let us consider two $messages$ $M_1$ and $M_2$ which when encrypted, produce ciphertexts $C_1$ and $C_2$. In the event of $key$ and $nonce$ being reused, in a deterministic stream cipher, $C_1 \oplus C_2 = M_1 \oplus M_2$. Whereas in Freestyle, for $|M_1|$ and $|M_2|$ $\geq log_2(N_c)$:

\begin{align}\label{eq:probmc}
Pr (C_1 \oplus C_2 = M_1 \oplus M_2) = \frac{1}{N_c}
\end{align}
The equation \ref{eq:probmc} indicates that Freestyle is resistant to key re-installation attacks like KRACK \cite{vanhoef2017key}. Also, in existing approaches of ciphertext randomization, in case of $key$ and $nonce$ being reused, the random bytes to be shared with receiver are prone to XOR attacks. However, this is not possible with Freestyle, as  only $hash$es are sent to the receiver. And the random bytes are never sent to the receiver neither in plain or encrypted form.

\subsection{\label{sec:resisting}Resisting brute-force and dictionary attacks}
Freestyle cipher can resist brute-force and dictionary attacks in \emph{three} ways: (i) By keeping the {$cipher\_parameter$} secret, (ii) Restricting pre-computation of stream, (iii) Wasting adversary's time and computational power.
\subsubsection{\label{sec:cipher-bf}By keeping {$cipher\_parameter$} a secret}
 In Freestyle cipher, the secrecy of the plaintext depends only on the secrecy of the $key$; and the $cipher\_parameter$ in general need not be kept secret. {The main purpose of {$cipher\_parameter$} (figure \ref{fig:config}) is to discourage reuse of cryptanalysis data collected from weaker $cipher\_parameter$}s. However, if kept secret, it can resist brute-force attacks. Assuming the adversary guesses that {$R_{min}$} and {$R_{max}$} values are in the range $[a, b]$, where {$a$} and {$b$} are divisible by $H_I$ and {$a \le b$}. Then, the total possible values of {$(R_{min},R_{max})$} adversary has to try is:

\begin{align}\label{eqxy}
\left(\displaystyle\frac{b-a}{H_I} + 1\right) + \left(\displaystyle\frac{b-a}{H_I}\right)  + \left(\displaystyle\frac{b-a}{H_I} - 1\right)  + .... + 1\\\nonumber\\
\text{or}\qquad\displaystyle\frac{\displaystyle\left(\frac{b-a}{H_I}+ 1\right)\displaystyle\left(\frac{b-a}{H_I} + 2\right)}{2}
\end{align}

As $H_C \in \{1,2,3\}$ the number of possible values of {$(R_{min},R_{max},H_C)$} the adversary has to try is:
\begin{align} \label{eqxy2}
\displaystyle\frac{3}{2}\times\left(\frac{b-a}{H_I}+ 1\right)\displaystyle\left(\frac{b-a}{H_I} + 2\right)
\end{align}
If the adversary's guesses for $R_{min}$, $R_{max}$ is represented as $R^*_{min}$ and $R^*_{max}$, then:
\begin{align} \label{pp}
R^*_{min} =&\left\{a, a+1, .. , b\right\}\\
R^*_{max} =&\{a,a+1, ..., b\}
\end{align}
Such that the guessed $R_{min} \le R_{max}$, and the value of $H_I\in c_f(R_{min},R_{max})$, where $c_f (R_{min},R_{max})$ is a set containing common factors of $R_{min}$ and $R_{max}$. Also, as $I_C \in [12,36]$ there are 25 possible values of $I_C$. Then, the total possible values of {$(R_{min},R_{max},H_C, H_I, I_C)$} or $cipher\_parameter$s are:

\begin{align} \label{eqxy3}
\sum_{n \in R^*_{min}}\:\sum_{m \in R^*_{max}\atop m \ge n}\:\displaystyle\sum_{h \in c_f(m,n)}\frac{75}{2}\displaystyle\left(\frac{m-n}{h}+ 1\right)\displaystyle\left(\frac{m-n}{h} + 2\right)
\end{align}
For example, if an adversary guesses $a=8$ and $b=32$, then using the equation \ref{eqxy3}, the effort required for the brute-force attack increases by {42525 ($\approx 2^{15}$}) times. Thus, for an effective attack, the adversary has to depend on other sources of information to guess the {$cipher\_parameter$}.

\subsubsection{Restricting pre-computation of key-stream}
In ChaCha, the key-stream can be pre-computed for various $key$s if $nonce$ is known. Pre-computation of stream is advantageous for a genuine receiver, as there is no need to wait for the data. However, for an adversary, pre-computation of streams with various $key$s is ideal to perform brute-force and dictionary attacks.

In Freestyle, since the key-stream depends on the $random\_words$ and {$hash$}, the exact key-stream cannot be pre-computed unless the sender sends expected $hash$es. This however, also restricts pre-computation of key-stream for a genuine receiver.

\subsubsection{\label{sec:wasting-adversarys-time}Wasting adversary's time and computational resources}
Here we introduce the Key-guessing penalty (KGP) metric which indicates the penalty an adversary has to pay in terms of computational power if an incorrect $key$ is used.
\\

\stepcounter{def}
\noindent{\bf Definition : Key-guessing penalty (KGP)} - The ratio of expected time taken for attempting to decrypt a $message$ using an incorrect $key$ and the expected time taken to decrypt a $message$ using a correct $key$ (equation \ref{eq:penaltygeneral}).
\begin{equation}\label{eq:penaltygeneral}
\begin{split}
&\displaystyle\frac{T(\text{attempt to decrypt a $message$ using a {incorrect} $key$})}{T(\text{decrypt a $message$ using the {correct} $key$})}
\end{split}\end{equation}
KGP is the measure of a cipher's resistance to brute-force and dictionary attacks. Based on KGP, a cipher can be classified in to two categories (i) Ciphers with KGP $\le 1$, which are not resistant to brute-force and dictionary attacks; and (ii) KGP $> 1$, ciphers that are brute-force and dictionary attack resistant. Ciphers with KGP $> 1$ are useful in scenarios where an adversary has higher computational power (e.g. a high end laptop) than the attacked system (e.g. a low powered IoT device). Such ciphers forces the adversary to use a machine that is at least KGP times faster than the attacked system, to launch an effective attack.\\

\stepcounter{remark}
\noindent{\bf Remark \arabic{remark} } 
{While computing KGP, the probability that an adversary can detect if the guessed $key$ is incorrect must be taken into account.} For ciphers with KGP $> 1$, for a given length of $message$, the amount of time required by an adversary to detect if the attempted $key$ is incorrect must be greater than the time taken to attempt decryption using the incorrect $key$.

\stepcounter{remark}
\noindent{\bf Remark \arabic{remark} } KGP $>1$ may also be achieved by using delays and CAPTCHAs for each incorrect $key$ attempt. However, this this not due to the property of the cipher itself. Also, such techniques are not useful in resisting offline brute-force and dictionary attacks.\hfill$\blacksquare$

If the sender uses uniform distribution to select the $pepper$ value, the $E_{pepper}$ will be $2^{(I_C -1)}$; however, for an adversary, since the $hash$es are unlikely to match, would require $2^{I_C}$ attempts. Hence, the maximum KGP one can expect using uniform distribution is $\approx$ 2. To improve KGP, the sender must use a right-skewed distribution which is kept secret and is not needed to be shared with the receiver. A right-skewed distribution is the one which tends to use smaller values for $pepper$.

\stepcounter{remark}
\noindent{\bf Remark \arabic{remark} } Irrespective of the distribution used to generate $pepper$ and the number of rounds for encryption/decryption, to generate $random\_word$s a good (pseudo-)random number generator with uniform distribution must be used.\hfill$\blacksquare$

As mentioned earlier in section \ref{sec:random-init}, during initialization a temporary configuration of $R_{min} = 12, R_{max} = 36, H_I = 1, H_C = 3$ is set. When an adversary uses an incorrect $key$, the probability of having a collision for a 16-bit $hash$ changes in each trial, and not all $hash$es have equal probability of occurring. Also, it must be noted that $hash$es are picked without replacement i.e. if a collision occurs, the $hash$ is incremented until there is no collision. Then, in the worst-case scenario, the maximum difference between the probability of getting two $hash$es which may occur at the 24$^{th}$ trial is:
\begin{align}
 \left(\frac{25}{2^{16} - 24}\right) - \left(\frac{1}{2^{16} - 24}\right) = 0.0003
\end{align}
which is negligible value for all practical purposes. Hence, for simplicity, we present approximate results assuming that all the $hash$es at a given trial are equally likely. Then, the probability of colliding a 16-bit $hash$ at the $n^{th}$ trial when an incorrect $key$ or $pepper$ is used (denoted by $Pr_n(X=1)$) is given as:
\begin{align}
\begin{cases}
  \displaystyle\frac{1}{2^{16}},& \text{if}\:\:\: n = 1\\
  &\\
\displaystyle\left(\frac{1}{2^{16} - n + 1}\right)\times\displaystyle\prod_{i=0}^{n-1}\displaystyle\left(\frac{2^{16}-i-1}{2^{16}-i}\right),&\text{Otherwise}
\end{cases}
\end{align}
\begin{figure*}
 \line(1,0){515}
 \begin{align}\label{eq:big}
E_{R_w} &\approx \displaystyle\sum_{h=1}^{28}\left(\displaystyle\sum_{n=1}^{N_r}Pr_n(X=1)\right)^{h-1}\left[{\left(\displaystyle\sum_{n=1}^{N_r}\left(R_{min} + n H_I\right) Pr_n(X=1)\right)} + {R_{max} \left(1 - \displaystyle\displaystyle\sum_{n=1}^{N_r}Pr_n(X=1)\right)}\right] \approx 36.0095
\end{align}
 \line(1,0){515}
\end{figure*}

Then, the expected number of rounds a user with an incorrect $key$ or $pepper$ will execute is denoted by $E_{R_w}$ can be computed as given in equation \ref{eq:big}, i.e. $E_{R_w} \approx 36.0095$.

During the cipher initialization, for a correct $key$ and $pepper$, the expected number of rounds a user will execute is 24 (i.e. average of 12 and 36). After initialization, $R_{min}$, $R_{max}$, and $H_C$ are set to their original values, and while decryption, if the expected number of rounds a genuine user executes is denoted by $E_R$. To compute KGP using equation \ref{eq:penaltygeneral}, the adversary has to execute $2^{I_C} \times E_{R_w}$ rounds during initialization, and $E_R$ rounds to decrypt a single block of $message$. Where as a genuine user has to run $E_{pepper} \times E_{R_w}$ rounds during initialization, and $28 \times 24$ rounds when using the correct $pepper$, and $N_b \times E_R$ rounds to decrypt a $message$ of $N_b$ blocks. Hence KGP is computed as:

\begin{align}\label{eq:penalty}
\text{KGP} &=&\displaystyle\frac{2^{I_C} \times E_{R_w} + E_R \times{\left(\displaystyle\sum_{n=1}^{N_r}Pr_n(X=1)\right)^{28}}}{E_{pepper}\times E_{R_w} + 28\times 24+ N_b\times E_{R}}
\end{align}\\

The probability of getting all the 28 $hash$es correct and attempting to decrypt the first block of $message$ using an incorrect $key$ is:
\begin{align}
\left(\displaystyle\sum_{n=1}^{N_r}Pr_n(X=1)\right)^{28} \approx 10^{-96}
\end{align}

which is negligible for all practical purposes. Hence:

\begin{align}\label{eq:penalty1}
\text{KGP}&\approx&\displaystyle\frac{2^{I_C} \times E_{R_w}}{E_{pepper}\times E_{R_w} + 28\times 24+ N_b\times E_{R}}
\end{align}

{i.e. for KGP $>$ 1:}
\begin{equation}
\begin{array}{ccc}
E_{pepper}  &< &2^{I_C} -  \left(\displaystyle\frac{672 + N_b\times E_{R}}{36.0095}\right)
\end{array}
\end{equation}

\begin{figure*}
\centering
\begin{tabular}{cc}
\includegraphics[scale=0.6]{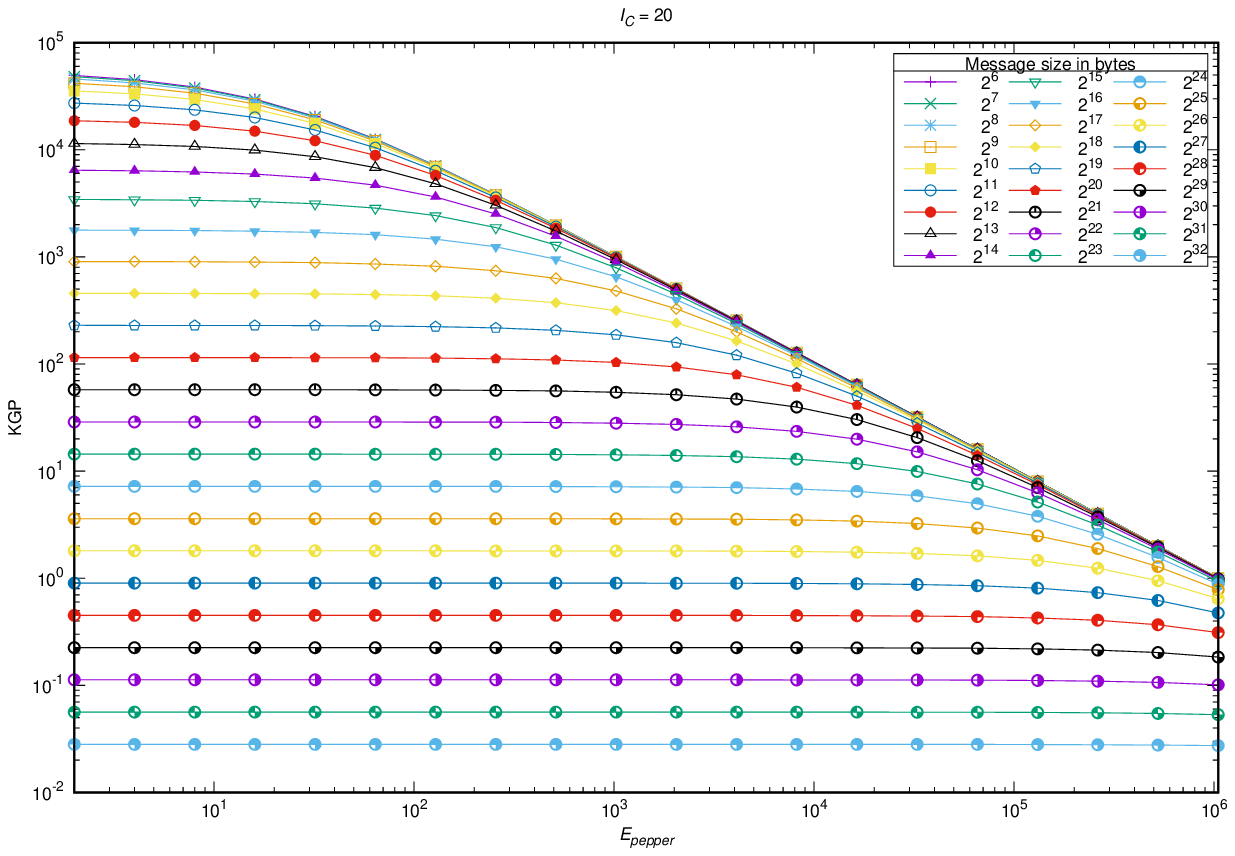}&\includegraphics[scale=0.6]{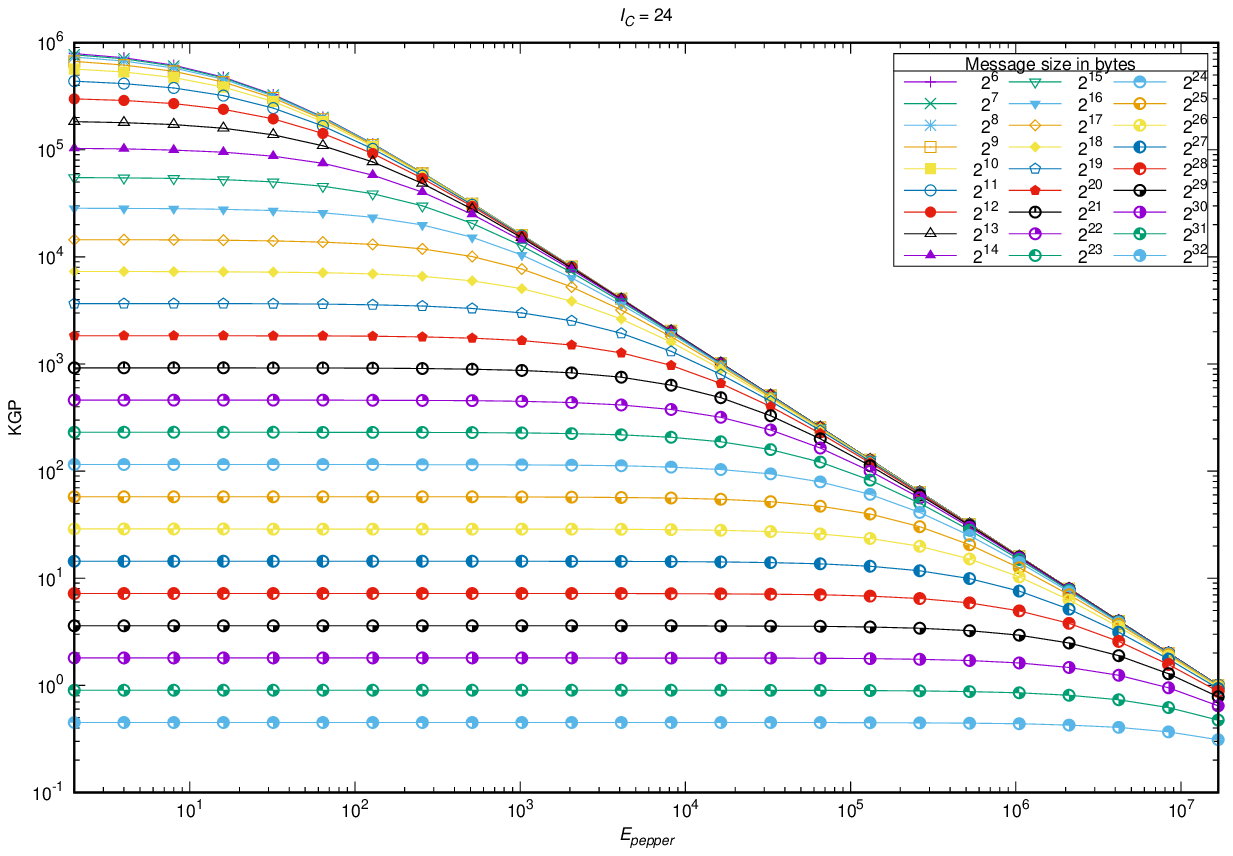}\\
\includegraphics[scale=0.6]{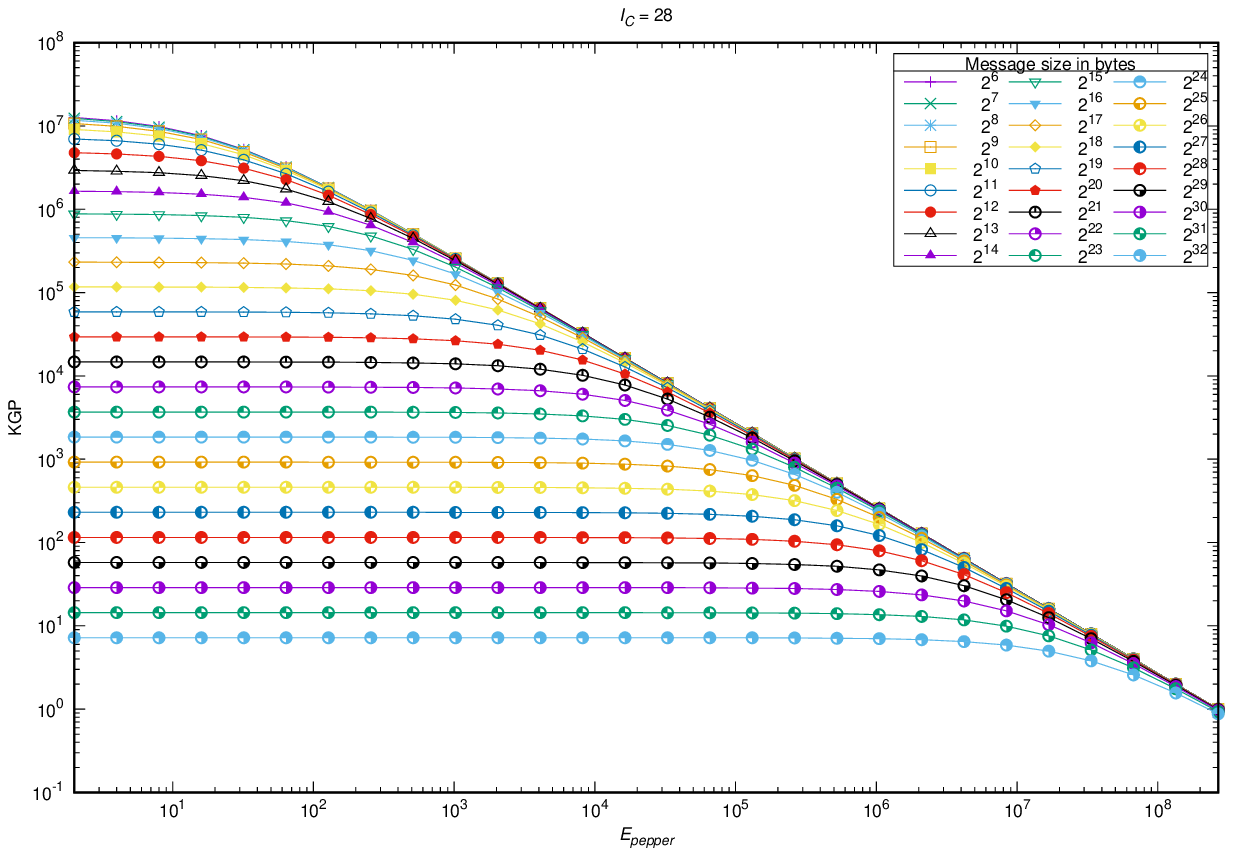}&\includegraphics[scale=0.6]{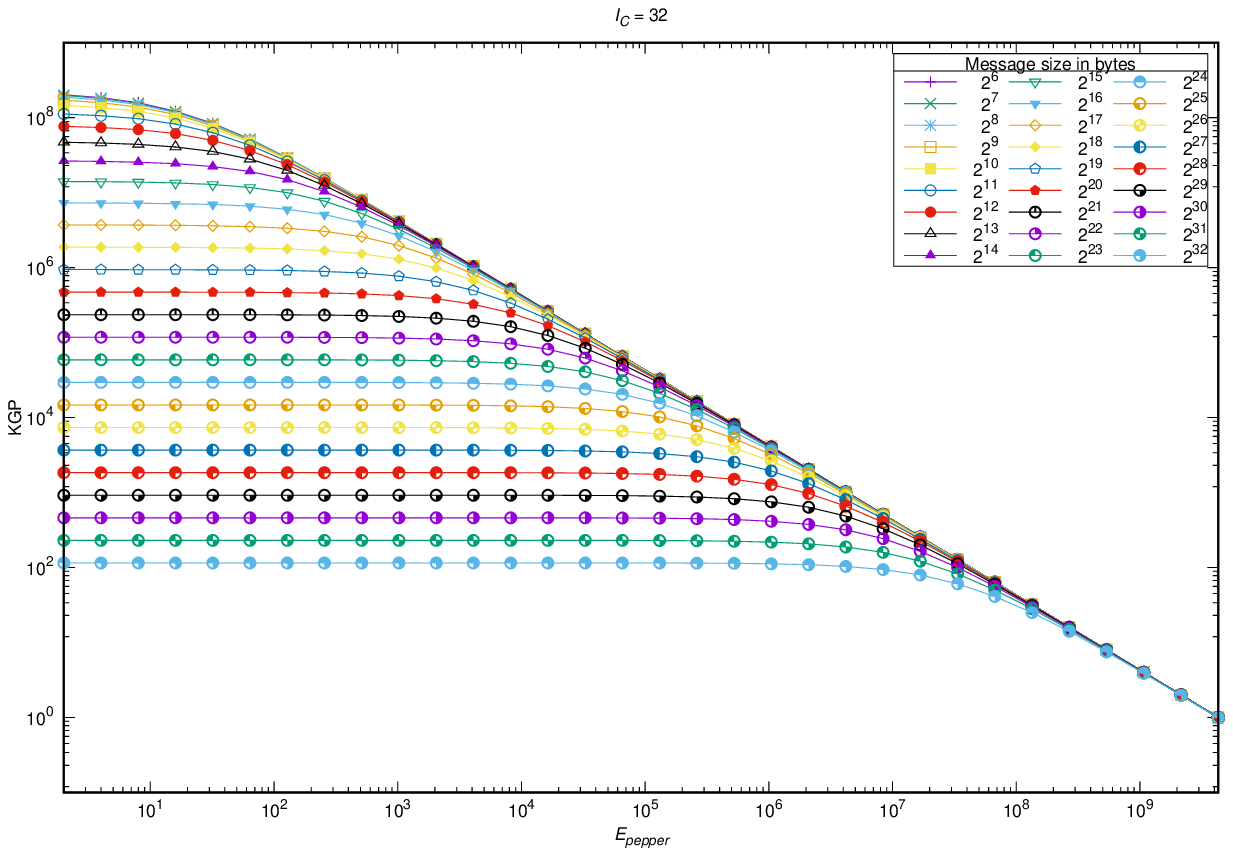}
  
\end{tabular}
\caption{\label{fig:kgp}KGP vs $E_{pepper}$ for $R_{min} = 8, R_{max} = 32, H_C = 3, H_I = 1$, $E_R = 20$, $I_C \in \{20, 24, 28, 32\}$, and various message sizes (64 bytes to 4GB)}
\end{figure*}

The value of $E_{pepper}, E_R,$ and $I_C$ can be chosen considering the performance, security level, and the required KGP. The figure \ref{fig:kgp} shows the result of KGP vs. $E_{pepper}$ for $R_{min} = 8, R_{max} = 32, H_C = 3, H_I = 1$, $E_R = 20$, $I_C \in \{20, 24, 28, 32\}$, and various message sizes 64 bytes to 4GB.

\subsection{\label{sec:128}Better security for 128-bit keys}
Though, not recommended, ChaCha supports 128-bit $key$s by concatenating the $key$ with itself to form a 256-bit $key$. In Freestyle, $cipher\_parameter$ and $random\_word$s are used to modify the initial state of cipher to provide an additional 128-bit random secret (in the form of $random\_word$s). The $random\_word$s are statistically independent of the $key$ and $nonce$ (equation \ref{eq:initial-state}); hence, for applications where 128-bit $key$s have to be used, Freestyle offers better security than ChaCha.

\subsection{\label{sec:overhead}Overheads}

\subsubsection{Computational overhead}
Freestyle has two main overheads when compared to ChaCha: (i) Overhead in generating a random number for each block of $message$; (ii) Computation of a hash after every $H_I$ rounds, which uses $H_C+2$ quarter rounds of Freestyle. Hence, the computational overhead for encryption is:
\begin{equation} \label{eq3}
\begin{array}{lll}
= T(\text{generate $N_b$ random numbers})\\
+\:\displaystyle\sum_{i=1}^{N_b} \left(\frac{R_i - R_{min}}{H_I} + 1\right) \left(H_C + 2\right)\times T (\text{1 QR of Freestyle})
\end{array}
\end{equation}

\begin{figure}
\centering
\includegraphics[scale=0.72]{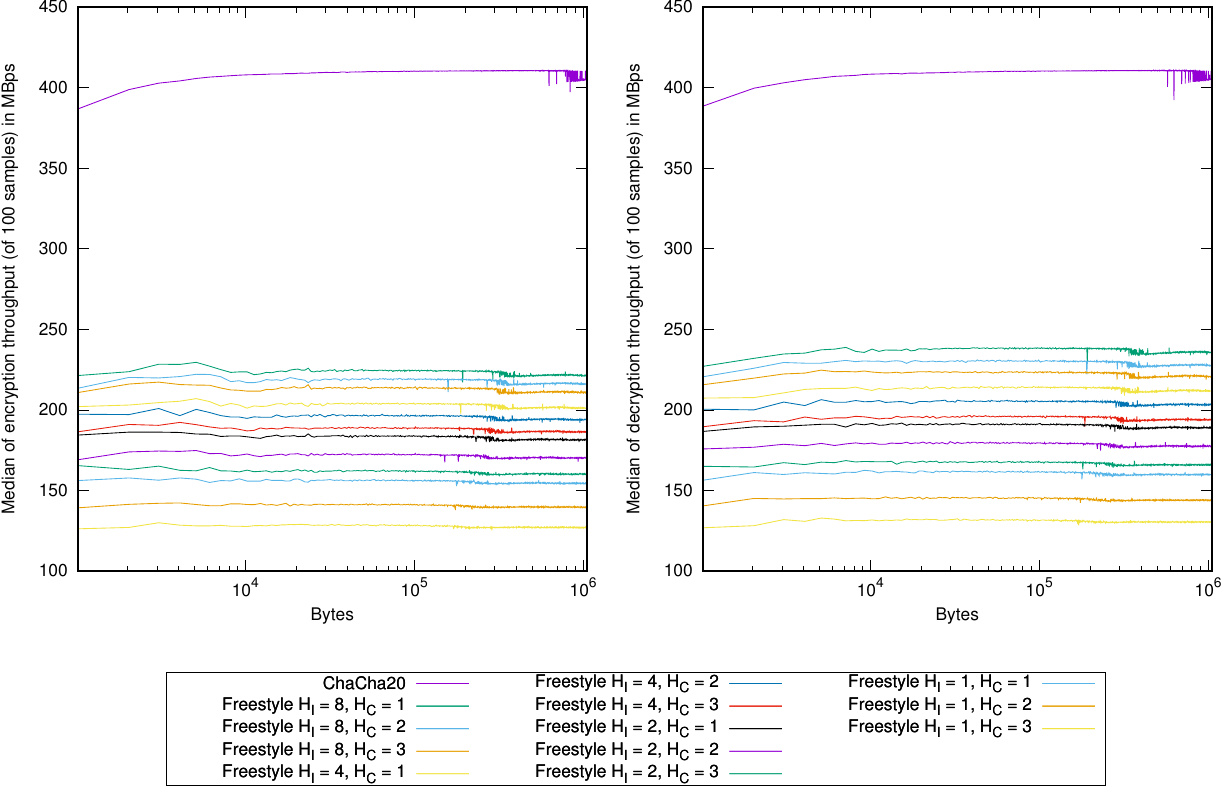}
\caption{\label{fig:perf}Performance comparison of Freestyle with $R_{min} = 8$, $R_{max} = 32$  \emph{vs} ChaCha20 on Intel Core i5-6300HQ processor without accounting for the time taken for cipher initialization}
\end{figure}

The worst case performance overhead is when $R_i = R_{max}, \forall i$. The figure \ref{fig:perf} shows the comparison of performance between optimized versions of and ChaCha20\footnote{\url{http://cvsweb.openbsd.org/cgi-bin/cvsweb/src/usr.bin/ssh/chacha.c?rev=1.1}} and Freestyle\footnote{\url{https://github.com/arun-babu/freestyle}} with various configurations without accounting for the time taken for initialization. The results were obtained on Intel Core i5-6300HQ processor with {\tt arc4random}\cite{arc4random} as the random number generator on OpenBSD. For the performance tests, $R_{min}=8$, $R_{max}=32$ has been used to make the cipher performance comparable to ChaCha20, as an uniformly distributed random number generator is used. The results indicate that Freestyle could be 1.6 to 3.2 times slower than ChaCha20 (figure \ref{fig:perf}).

\subsubsection{Bandwidth overhead}
Freestyle algorithm requires a sender to send the final round 16-bit $hash$; i.e. requiring to send send extra 8 bits for each block of $message$ to be sent. Also, for initialization of $random\_word$, it requires extra $28\times 16$ bits. Hence, the total bandwidth overhead in bits is $16N_b + 28\times 16$, i.e.
\begin{equation} \label{eq4}
\begin{split}
\text{Bandwidth overhead (in \%)}&= \frac{16N_b + 512}{|message|}\times 100
\end{split}
\end{equation}
For a $message$ of length in multiples of 512 bits, the bandwidth overhead is $\approx$ 3.125\%.

\subsection{Side-channel attacks}
Freestyle uses Add-Rotate-XOR instructions to resist timing and side channel attacks. However, if a device leaks information related to randomness, Freestyle is at least as secure as ChaCha with $R_{min}$ rounds. This is because, from $hash$ at least $2^{256}$ operations are required to generate possible cipher states ($S^{(R)}$), and since $hash$ is a 16 bit value, there are likely to be $2^{248}$ cipher states that collide a given $hash$. For devices that have weak or insecure (pseudo-)random number generator, we recommend a conservative configuration of $H_C = 3$; in which case at least $2^{512}$ operations are required to generate the cipher states colliding a given $hash$.

Another potential attack in Freestyle could be to gather intermediate $hash$es. While performing encryption, Freestyle requires maintaining the hash collision information for various {$hash$}es. One of the simplest implementation is to use a look-up table implementations which are prone to timing attacks. Although, by leaking intermediate $hash$es it is computationally infeasible to compute the $key$ or key-stream. However, with the leaked $hash$, the adversary can detect if the attempted $key$ is correct or not after $R_{min}$ rounds; making KGP $< 1$. Also, some cryptanalysis advantage may be gained by observing consecutive $hash$es. Although, this would also require knowledge of $cipher\_parameter$ and 128-bit $random\_word$s.

Such attacks can be resisted by obscuring look-up table indices by XOR-ing it with a 16-bit random mask. The random mask is to be generated for each block of $message$ to be encrypted/decrypted and provides $2^{16}$ different timing variations. Though, such techniques may resist timing attacks; but do not guarantee protection against such attacks. Also, computation and memory overheads must be taken into account before considering such implementations.

Observing the time taken for encryption/decryption may be used by an adversary to predict the $pepper$ and thus reducing KGP. To prevent such attacks, the $pepper$ value can be shared with the receiver through a secure channel. However, this approach is equivalent to increasing the $key$ size by $I_C$ bits.
\section{\label{sec:related-work}Related work}
 \subsection{\label{sec:randomized-ciphers}Randomized encryption schemes}
Use of randomized encryption schemes have been in practice for many years, and a taxonomy of randomized ciphers is presented in \cite{rivest1983randomized}. Also, some approaches to randomized encryption for public-key cryptography was proposed in \cite{goldwasser1984probabilistic,elgamal1985public,cramer1998practical}.
Approaches based on chaotic systems for probabilistic encryption were also proposed \cite{papadimitriou2001probabilistic}. However, the main concern with some of the existing approaches are high bandwidth expansion factor and computational overhead \cite{li2003problems,rivest1983randomized}.

The key difference between existing approaches and the current work is: the random bytes are never sent to the receiver in plain nor in the encrypted form. The random bytes are to be computed by the receiver from the initial 28 $hash$es. The initial 28  $hash$es also serve the purpose of preventing an adversary from sending arbitrary ciphertext, thus resisting CCA if the $nonce$ cannot be controlled by the adversary. Also, Freestyle offers the possibility of generating $2^{128}$ different ciphertexts even if $key$, $nonce$, and other cipher parameters are reused. Also unlike some of the existing randomized ciphers, Freestyle has a low bandwidth overhead of $\approx$ 3.125\%.
 \subsection{\label{sec:difficulty-approaches}Approaches based on difficulty and proof of work}
Several algorithms have been proposed in literature to increase the difficulty in key and password guessing using an CPU intensive key-streaching\cite{kelsey1997secure} or key-setup phase \cite{provos1999bcrypt} using a cost-factor. Also approaches that consume large amount of memory have also been proposed \cite{percival2016scrypt,forler2013catena}. Another related area is use of client puzzles \cite{boyen2007halting} and proof-of-work (e.g. Bitcoin \cite{nakamoto2008bitcoin}) to delay cryptographic operations. 

The \emph{hash based halting condition} described in section \ref{sec:hbsc}, on a high-level uses similar principle as the Halting key derivation function (HKDF) proposed in \cite{boyen2007halting}. In HKDF, a sender using a password and a random bytes uses the key derivation function till $n$ iterations (or based on certain amount of time) to generate a $key$ and a publicly verifiable $hash$. On the other hand, the receiver uses the random bytes and password to generate the $key$ till the verifiable $hash$ matches.

Our approach however differs from \cite{boyen2007halting} in the following ways: (i) The minimum and maximum number of iterations is explicitly defined and is expected to be public. This step is crucial as it ensures a minimum level of security for genuine user during encryption/decryption. It also ensures that an adversary executes at least the minimum number of iterations.  The maximum iterations ensures that a genuine user cannot run more than specified iterations; thus preventing the possibility of DoS attacks or getting stuck in an infinite loop due to human errors;
 (ii) Freestyle does not require a complex collision resistant hash function, as $hash$ collisions are handled simply incrementing the $hash$ if a collision occurs. Also, the hash function uses ARX instructions to resist any side-channel cryptanalysis; (iii) In Freestyle, the security of the cipher is not dependent on amount of time taken or number of iterations for cipher initialization, but on the length of $pepper$ bits; (iv) Freestyle uses a 28 number of 16-bit $hash$es for initialization and a 16-bit $hash$ for every block of message being sent, thus the total size of $hash$ is not fixed and is $\propto |message|$; (v) Freestyle does not require $hash$ computation at every iteration, instead a hash interval ($H_I$) parameter is used to determine round intervals at which $hash$ must be computed, thus offering flexibility to adjust performance and security. Similarly, Freestyle offers flexibility in choosing the complexity of hash function using a hash complexity ($H_C$) parameter; and (vi) Freestyle forces the cipher initialization with $R_{min} = 12$ and $R_{max} = 36$, thus ensures enough randomness even in cases where user provides insecure parameters for cipher initialization; and (vi) Freestyle offers the possibility of much higher KGP by allowing the sender to choose a right-skewed distribution to generate $pepper$ and $R_i$.
 
 \subsection{\label{sec:critical-review}Freestyle \emph{vs} ChaCha}
When compared to ChaCha, Freestyle offers better security for 128-bit keys (section \ref{sec:128}). It also provides the possibility of generating $2^{128}$ ciphertexts for a given $message$ even if $nonce$ and $key$ is reused (section \ref{sec:num-possible-ciphers}). This makes Freestyle resistant to XOR of ciphertext attacks if $key$ and $nonce$ is reused. Randomization also makes Freestyle resistant to KPA, CPA, and CCA (section \ref{sec:cryptanalysis}). Freestyle offers the possibility of KGP $> 1$, which makes it resistant to brute-force and dictionary based attacks (section \ref{sec:resisting}).

On the other hand, Freestyle is {1.6 to 3.2 times slower} than ChaCha (section \ref{sec:overhead}), and also has a higher cost of initialization (section \ref{sec:random-init}). In terms of bandwidth overhead, Freestyle generates $\approx$ 3.125\% larger ciphertext. And, in implementation overhead, Freestyle's encryption and decryption logic differ slightly. ChaCha is a simple constant time algorithm, where as Freestyle is a randomized algorithm, and assumes that the sender has a good source of random numbers.

\section{\label{sec:conclusion}Conclusion}
In this paper we have introduced Freestyle, a novel randomized cipher capable of generating $2^{128}$ different ciphertexts for a given $key$, $nonce$, and $message$; making known-plaintext (KPA), chosen-plaintext(CPA) and chosen-ciphertext(CCA) attacks difficult in practice. We have introduced the concepts of bounded \emph{hash based halting condition} and \emph{key-guessing penalty} (KGP), which are helpful in development and analysis of ciphers resistant to key-guessing attacks. Freestyle has demonstrated KGP $> 1$ which makes it run faster on a low-powered machine having the correct $key$, and is KGP times slower (with high probability) on an adversary's machine. Freestyle is ideal for applications where the ciphertext is assumed to be in full control of the adversary i.e. where an offline brute-force or dictionary attack can be carried out. Example use-cases include
disk encryption, encrypted databases, password managers, sensitive data in public facing IoT devices, etc.
The paper has introduced a new class of ciphers having KGP $> 1$. There is further scope for research on other possible and simpler ways to achieve KGP $> 1$, and study the properties of such ciphers. The possibility of forcing an adversary to solve a NP-hard problem for every decryption attempt with an incorrect $key$ could be an attractive topic of research. The key challenge however is to make the time taken for decryption attempt with an incorrect $key$, greater than the time taken to detect if the problem is NP-hard.

\ifCLASSOPTIONcaptionsoff
  \newpage
\fi



%

\section*{Acknowledgments}
This work was supported in part by the Bosch Research and Technology Centre - India under the project titled "E-sense - Sensing and Analytics for Energy Aware Smart Campus", and in part by the Robert Bosch Centre for Cyber-Physical Systems, Indian Institute of Science, Bengaluru. The authors thank Sagar Gubbi, Rajesh Sundaresan, Navin Kashyap, and Sanjit Chatterjee for helpful discussions.


\begin{thebibliography}{10}

\bibitem{bernstein2008chacha}
D.~J. Bernstein, ``{ChaCha, a variant of Salsa20},'' in {\em Workshop Record of
  SASC}, vol.~8, pp.~3--5, 2008.

\bibitem{langley2016chacha20}
A.~Langley, W.~Chang, N.~Mavrogiannopoulos, J.~Strombergson, and S.~Josefsson,
  ``Chacha20-poly1305 cipher suites for transport layer security (tls),'' tech.
  rep., 2016.

\bibitem{djm2}
D.~Miller and S.~Josefsson, ``{The chacha20-poly1305@openssh.com authenticated
  encryption cipher draft-josefsson-ssh-chacha20-poly1305-openssh-00}.''
\newblock Network Working Group Internet-Draft,
  \url{https://tools.ietf.org/html/draft-josefsson-ssh-chacha20-poly1305-openssh-00},
  Last accessed 1.12.2018.

\bibitem{djm}
D.~Miller, ``chacha20poly1305 protocol.''
\newblock
  \url{https://cvsweb.openbsd.org/cgi-bin/cvsweb/src/usr.bin/ssh/PROTOCOL.chacha20poly1305?annotate=HEAD},
  Last accessed 1.12.2018.

\bibitem{chacha-perf}
``{eBACS: ECRYPT Benchmarking of Cryptographic Systems}.''
\newblock \url{https://bench.cr.yp.to/results-stream.html}.

\bibitem{chacha-google-perf}
E.~Bursztein, ``{Speeding up and strengthening HTTPS connections for Chrome on
  Android},'' Apr. 2014.
\newblock Google security blog,
  \url{https://security.googleblog.com/2014/04/speeding-up-and-strengthening-https.html}.

\bibitem{cve-infeon}
``{Vulnerability Note VU\#307015, Infineon RSA library does not properly
  generate RSA key pairs},'' Oct 2017.
\newblock CVE-2017-15361, \url{https://www.kb.cert.org/vuls/id/307015}.

\bibitem{kim2013predictability}
S.~H. Kim, D.~Han, and D.~H. Lee, ``Predictability of android openssl's pseudo
  random number generator,'' in {\em Proceedings of the 2013 ACM SIGSAC
  conference on Computer \& communications security}, pp.~659--668, ACM, 2013.

\bibitem{bello2008predictable}
L.~Bello, M.~Bertacchini, and B.~Hat, ``Predictable prng in the vulnerable
  debian openssl package: the what and the how,'' in {\em the 2nd DEF CON
  Hacking Conference}, 2008.

\bibitem{yilek2009private}
S.~Yilek, E.~Rescorla, H.~Shacham, B.~Enright, and S.~Savage, ``When private
  keys are public: Results from the 2008 debian openssl vulnerability,'' in
  {\em Proceedings of the 9th ACM SIGCOMM conference on Internet measurement
  conference}, pp.~15--27, ACM, 2009.

\bibitem{lenstra2012ron}
A.~Lenstra, J.~P. Hughes, M.~Augier, J.~W. Bos, T.~Kleinjung, and C.~Wachter,
  ``Ron was wrong, whit is right,'' tech. rep., IACR, 2012.

\bibitem{heninger2012mining}
N.~Heninger, Z.~Durumeric, E.~Wustrow, and J.~A. Halderman, ``Mining your ps
  and qs: Detection of widespread weak keys in network devices.,'' in {\em
  USENIX Security Symposium}, vol.~8, 2012.

\bibitem{vanhoef2017key}
M.~Vanhoef and F.~Piessens, ``Key reinstallation attacks: Forcing nonce reuse
  in wpa2,'' in {\em Proceedings of the 24th ACM Conference on Computer and
  Communications Security (CCS)}, ACM, 2017.

\bibitem{beurdouche2015messy}
B.~Beurdouche, K.~Bhargavan, A.~Delignat-Lavaud, C.~Fournet, M.~Kohlweiss,
  A.~Pironti, P.-Y. Strub, and J.~K. Zinzindohoue, ``A messy state of the
  union: Taming the composite state machines of tls,'' in {\em Security and
  Privacy (SP), 2015 IEEE Symposium on}, pp.~535--552, IEEE, 2015.

\bibitem{adrian2015imperfect}
D.~Adrian, K.~Bhargavan, Z.~Durumeric, P.~Gaudry, M.~Green, J.~A. Halderman,
  N.~Heninger, D.~Springall, E.~Thom{\'e}, L.~Valenta, {\em et~al.},
  ``Imperfect forward secrecy: How diffie-hellman fails in practice,'' in {\em
  Proceedings of the 22nd ACM SIGSAC Conference on Computer and Communications
  Security}, pp.~5--17, ACM, 2015.

\bibitem{gu2017attacking}
W.~Gu, Y.~Huang, R.~Qian, Z.~Liu, and R.~Gu, ``Attacking crypto-1 cipher based
  on parallel computing using gpu,'' in {\em International Conference on
  Applications and Techniques in Cyber Security and Intelligence},
  pp.~293--303, Springer, 2017.

\bibitem{agosta2013high}
G.~Agosta, A.~Barenghi, and G.~Pelosi, ``High speed cipher cracking: the case
  of keeloq on cuda,'' 2013.

\bibitem{chiriaco2017finding}
V.~Chiriaco, A.~Franzen, R.~Thayil, and X.~Zhang, ``Finding partial hash
  collisions by brute force parallel programming,'' in {\em Systems,
  Applications and Technology Conference (LISAT), 2017 IEEE Long Island},
  pp.~1--6, IEEE, 2017.

\bibitem{wiemer2014high}
F.~Wiemer and R.~Zimmermann, ``High-speed implementation of bcrypt password
  search using special-purpose hardware,'' in {\em ReConFigurable Computing and
  FPGAs (ReConFig), 2014 International Conference on}, pp.~1--6, IEEE, 2014.

\bibitem{malvoni2014your}
K.~Malvoni, D.~Solar, and J.~Knezovi{\'c}, ``Are your passwords safe:
  Energy-efficient bcrypt cracking with low-cost parallel hardware,'' in {\em
  WOOT'14 8th Usenix Workshop on Offensive Technologies Proceedings 23rd USENIX
  Security Symposium}, 2014.

\bibitem{liu2017elliptic}
Z.~Liu, J.~Gro{\ss}sch{\"a}dl, Z.~Hu, K.~J{\"a}rvinen, H.~Wang, and
  I.~Verbauwhede, ``Elliptic curve cryptography with efficiently computable
  endomorphisms and its hardware implementations for the internet of things,''
  {\em IEEE Transactions on Computers}, vol.~66, no.~5, pp.~773--785, 2017.

\bibitem{javeed2016high}
K.~Javeed, X.~Wang, and M.~Scott, ``High performance hardware support for
  elliptic curve cryptography over general prime field,'' {\em Microprocessors
  and Microsystems}, 2016.

\bibitem{khalid2017rc4}
A.~Khalid, G.~Paul, and A.~Chattopadhyay, ``Rc4-accsuite: A hardware
  acceleration suite for rc4-like stream ciphers,'' {\em IEEE Transactions on
  Very Large Scale Integration (VLSI) Systems}, vol.~25, no.~3, pp.~1072--1084,
  2017.

\bibitem{gurkaynak2017multi}
F.~K. G{\"u}rkaynak, R.~Schilling, M.~Muehlberghuber, F.~Conti, S.~Mangard, and
  L.~Benini, ``Multi-core data analytics soc with a flexible 1.76 gbit/s
  aes-xts cryptographic accelerator in 65 nm cmos,'' in {\em Proceedings of the
  Fourth Workshop on Cryptography and Security in Computing Systems},
  pp.~19--24, ACM, 2017.

\bibitem{kim2016multistate}
W.~Kim, A.~Chattopadhyay, A.~Siemon, E.~Linn, R.~Waser, and V.~Rana,
  ``Multistate memristive tantalum oxide devices for ternary arithmetic,'' {\em
  Scientific reports}, vol.~6, 2016.

\bibitem{jain2017computing}
S.~Jain, A.~Ranjan, K.~Roy, and A.~Raghunathan, ``Computing in memory with
  spin-transfer torque magnetic ram,'' {\em arXiv preprint arXiv:1703.02118},
  2017.

\bibitem{sebastian2017temporal}
A.~Sebastian, T.~Tuma, N.~Papandreou, M.~L. Gallo, L.~Kull, T.~Parnell, and
  E.~Eleftheriou, ``Temporal correlation detection using computational
  phase-change memory,'' {\em Nature Communications}, 2017.

\bibitem{slow-is-good}
W.~Buchanan, ``{When Slow Is Good - The Great Slowcoach: Bcrypt},'' July 2015.
\newblock
  \url{https://www.linkedin.com/pulse/when-slow-good-great-slowcoach-bcrypt-william-buchanan}.

\bibitem{bernstein2005salsa20}
D.~J. Bernstein, ``Salsa20 specification,'' {\em eSTREAM Project algorithm
  description, http://www. ecrypt. eu. org/stream/salsa20pf. html}, 2005.

\bibitem{bernstein2008salsa20}
D.~J. Bernstein, ``The salsa20 family of stream ciphers,'' {\em Lecture Notes
  in Computer Science}, vol.~4986, pp.~84--97, 2008.

\bibitem{nir2015chacha20}
Y.~Nir and A.~Langley, ``{ChaCha20 and Poly1305 for IETF Protocols},'' tech.
  rep., 2015.

\bibitem{denisxchacha20}
F.~Denis, ``The xchacha20-poly1305 construction.''
\newblock
  \url{https://download.libsodium.org/doc/secret-key_cryptography/xchacha20-poly1305_construction.html}.

\bibitem{libsodium-report}
``Libsodium v1.0.12 and v1.0.13 security assessment,'' tech. rep., 2017.
\newblock
  \url{https://www.privateinternetaccess.com/blog/wp-content/uploads/2017/08/libsodium.pdf}.

\bibitem{arc4random}
T.~De~Raadt, ``{arc4random - randomization for all occasions},'' 2014.

\bibitem{kedem1999bruteforceattackonuni}
G.~Kedem and Y.~Ishihara, ``{Brute force attack on UNIX passwords with SIMD
  computer},'' 1999.

\bibitem{forler2013catena}
C.~Forler, S.~Lucks, and J.~Wenzel, ``Catena: A memory-consuming
  password-scrambling framework,'' tech. rep., Citeseer, 2013.

\bibitem{ylonen2006secure}
T.~Ylonen and C.~Lonvick, ``The secure shell (ssh) transport layer protocol,
  rfc 4253,'' 2006.

\bibitem{bernstein2005poly1305}
D.~J. Bernstein, ``The poly1305-aes message-authentication code.,'' in {\em
  FSE}, vol.~3557, pp.~32--49, Springer, 2005.

\bibitem{rivest1983randomized}
R.~L. Rivest and A.~T. Sherman, ``{Randomized encryption techniques},'' in {\em
  Advances in Cryptology}, pp.~145--163, Springer, 1983.

\bibitem{goldwasser1984probabilistic}
S.~Goldwasser and S.~Micali, ``Probabilistic encryption,'' {\em Journal of
  computer and system sciences}, vol.~28, no.~2, pp.~270--299, 1984.

\bibitem{elgamal1985public}
T.~ElGamal, ``A public key cryptosystem and a signature scheme based on
  discrete logarithms,'' {\em IEEE transactions on information theory},
  vol.~31, no.~4, pp.~469--472, 1985.

\bibitem{cramer1998practical}
R.~Cramer and V.~Shoup, ``A practical public key cryptosystem provably secure
  against adaptive chosen ciphertext attack,'' in {\em Annual International
  Cryptology Conference}, pp.~13--25, Springer, 1998.

\bibitem{papadimitriou2001probabilistic}
S.~Papadimitriou, T.~Bountis, S.~Mavroudi, and A.~Bezerianos, ``A probabilistic
  symmetric encryption scheme for very fast secure communication based on
  chaotic systems of difference equations,'' {\em International Journal of
  Bifurcation and Chaos}, vol.~11, no.~12, pp.~3107--3115, 2001.

\bibitem{li2003problems}
S.~Li, X.~Mou, B.~L. Yang, Z.~Ji, and J.~Zhang, ``Problems with a probabilistic
  encryption scheme based on chaotic systems,'' {\em International Journal of
  Bifurcation and Chaos}, vol.~13, no.~10, pp.~3063--3077, 2003.

\bibitem{kelsey1997secure}
J.~Kelsey, B.~Schneier, C.~Hall, and D.~Wagner, ``Secure applications of
  low-entropy keys,'' in {\em International Workshop on Information Security},
  pp.~121--134, Springer, 1997.

\bibitem{provos1999bcrypt}
N.~Provos and D.~Mazieres, ``Bcrypt algorithm,'' USENIX, 1999.

\bibitem{percival2016scrypt}
C.~Percival and S.~Josefsson, ``The scrypt password-based key derivation
  function,'' tech. rep., 2016.

\bibitem{boyen2007halting}
X.~Boyen, ``Halting password puzzles,'' in {\em Proc. Usenix Security}, 2007.

\bibitem{nakamoto2008bitcoin}
S.~Nakamoto, ``Bitcoin: A peer-to-peer electronic cash system,'' 2008.

\end{thebibliography}

%

\relax\begin{IEEEbiographynophoto}{P. Arun Babu}
is currently a Member of Technical Staff at Robert Bosch Center for Cyber-physical Systems at the Indian Institute of Science, Bengaluru, India. Arun holds a Ph.D in Engineering Sciences from Indira Gandhi Centre for Atomic Research, Kalpakkam, India. His areas of research include cyber-security and software engineering.
\end{IEEEbiographynophoto}



\begin{IEEEbiographynophoto}{Jithin Jose Thomas}
is currently a Senior systems software engineer at the Trakray Innovations, Bengaluru, India. Jithin holds a B.Tech degree in Electronics and Communications Engineering from National Institute of Technology, Calicut, India. He was with the Department of Electrical Communication Engineering, Indian Institute of Science, Bengaluru at the time of this work. His areas of research include cryptography, wireless networks, and data analytics.
\end{IEEEbiographynophoto}



\end{document}